\documentclass[final,3p,times,12pt]{elsarticle}

\usepackage{amssymb}
\usepackage{amsthm}

\biboptions{sort&compress}

\usepackage{amsmath}
\usepackage{bm}
\usepackage{mathrsfs}
\usepackage{color}
\usepackage{algorithm}
\usepackage{algorithmic}
\usepackage{listings}
\usepackage[misc,geometry]{ifsym}
\usepackage{amssymb}
\usepackage{amsmath,amsfonts,upgreek}
\usepackage{mathrsfs}
\usepackage{bm}
\usepackage{color} 
\definecolor{mygreen}{RGB}{28,172,0}
\definecolor{mylilas}{RGB}{170,55,241}

\usepackage{epstopdf}
\usepackage{algorithm}
\usepackage{algorithmic}
\usepackage{ulem}
\usepackage{booktabs}
 \usepackage[utf8]{inputenc}
 \usepackage[overload]{empheq}
 \usepackage{cases}
\usepackage[table]{xcolor}
\usepackage{cancel}
\makeatletter
\newcommand{\ALOOP}[1]{\ALC@it\algorithmicloop\ #1%
  \begin{ALC@loop}}
\newcommand{\ENDALOOP}{\end{ALC@loop}\ALC@it\algorithmicendloop}

\newcommand{\lb}{\left(}
\newcommand{\rb}{\right)}

\newcommand{\sv}{\lb\begin{array}{ccccccccccccccccc}}
\newcommand{\ev}{\end{array}\rb}
\newcommand{\sV}{\begin{bmatrix}}
\newcommand{\eV}{\end{bmatrix}}

\usepackage[justification=centering]{caption}
\usepackage[font=small]{caption}
\usepackage{url}
\journal{}

\begin{document}
\begin{frontmatter}


\title{Two-scale data-driven design for heat manipulation}
\author{Daicong Da}
\author{{Wei Chen}\corref{cor1}}\ead{weichen@northwestern.edu}
\cortext[cor1]{Corresponding author}
\address{Department of Mechanical Engineering, Northwestern University, Evanston, IL 60208, USA}

\begin{abstract}

Data-driven methods have gained increasing attention in computational mechanics and design. This study investigates a two-scale data-driven design for thermal metamaterials with various functionalities. To address the complexity of multiscale design, the design variables are chosen as the components of the homogenized thermal conductivity matrix originating from the lower scale unit cells. Multiple macroscopic functionalities including thermal cloak, thermal concentrator, thermal rotator/inverter, and their combinations, are achieved using the developed approach. Sensitivity analysis is performed to determine the effect of each design variable on the desired functionalities, which is then incorporated into topology optimization. Geometric extraction demonstrates an excellent matching between the optimized homogenized conductivity and the extraction from the constructed database containing both architecture and property information. The designed heterostructures exhibit multiple thermal meta-functionalities that can be applied to a wide range of heat transfer fields from personal computers to aerospace engineering.

\end{abstract}

\begin{keyword}

Data-driven methods \sep Thermal metamaterials  \sep Design optimization \sep Homogenization \sep Heat manipulation  \sep Heat conduction
\end{keyword}

\end{frontmatter}

\section{Introduction} 

Data-driven methods have continuously received attention in computational mechanics to achieve various goals. Specifically, in terms of computational topology optimization for structures and materials, machine learning and data-driven algorithms have been applied to different scenarios with specific goals. First, because topological optimization often requires hundreds of design iterations, data-driven methods have sought to obtain an optimized structure or material without any iterations, (i.e., iteration-free optimization). Where optimization iterations cannot be avoided, data-driven methods have explored how to accelerate both the forward and inverse processes in every iteration, e.g., by using machine learning to accelerate or even replace the traditional finite element (FE) solver. Moreover, in order to update physical model at every iteration, topological design often requires gradient information at a specific objective function with respect to (w.r.t.) the design variables. In the cases where the gradient information is difficult or impossible to obtain, for certain engineering problems and extreme conditions, machine learning algorithms can be helpful to build the surrogate model between the input and output of interest. Finally, when it comes to multiscale design optimization, the design space is significantly expanded when the lower scale architectures are irregular, i.e., for non-periodic multiscale structures, which therefore requires novel data-driven design approaches to reduce the design dimensionality. The most recent review on the use of machine leaning on topological design optimization of mono-structure can be found in \cite{woldseth2022use}. A comprehensive review of database management, data-driven design of metamaterials, and (concurrent) multiscale design can be found in \cite{IDEALreview}.

For non-periodic heterostructures, where different local domains are populated with different building blocks, multiscale design by leveraging data-driven techniques and by directly tailoring the effective macroscopic property is an efficient way to reduce the design dimension. This strategy which will henceforth be referred to as \textit{property} design. This strategy relies heavily on the homogenization methodology to connect the two different scales, i.e., the low material unit cell or building block scale and the higher structure scale. In mono-scale problems for structural design only, the material constitutive relation is given, and the number of design variables is equal to the number of background FE mesh used in the current simulation. By contrast, for multiscale design, the material constitutive law originated from the topology (architecture) of the representative volume element (RVE) must be tailored. Directly updating topology via classical pseudo-density within RVE would be dimensionally very expensive, while the dimensionality of the effective properties, e.g., the homogenized elasticity or conductivity tensor, is often much smaller than the pixels or finite elements within RVE. For non-periodic multiscale structures where a large number of RVEs exist, the design dimension can be greatly reduced.

The first multiscale design work using the \textit{property} design strategy is reported in \cite{panetta2015elastic}, with the distinction of updating the pseudo density associated with every finite element at the material scale. This work uses only two \textit{property} parameters to represent one RVE because of the isotropic nature. The optimized property distributions are obtained by carrying out the topology optimization for targeting a given displacement at the structure scale. The final 3D non-periodic multiscale structure is assembled by searching the database to find the RVEs that match the optimized properties and filling in every structural finite element with one single substituted RVE. Because of the versatility of the above multiscale \textit{property} design methodology, the framework has been extended to orthotropic and anisotropic RVEs \cite{wang2020deep} for compliance minimization \cite{li2019design,zheng2021data,wang2022ih}, targeting a given deformation \cite{zhu2017two,wang2022ih}, stress minimization and fracture resistance \cite{da2022data}, as well as dynamic problems \cite{wang2022generalized}. A similar strategy is employed in this work for multiple thermal meta-functionalities, which is to the best of our knowledge the first time in literature that the data-driven design approach is used for such problems.

Thermal meta-functionalities including thermal cloak \cite{han2013homogeneous,schittny2013experiments,han2014full,xu2014ultrathin}, thermal concentrator \cite{guenneau2012transformation,chen2015experimental}, and thermal rotator/reversal \cite{vemuri2014guiding} have been long studied by using transformation theory \cite{leonhardt2006optical,pendry2006controlling, huang2019theoretical}, both numerically \cite{fan2008shaped,peralta2017optimization,zhou2019while} and experimentally \cite{narayana2012heat,dede2013heat, han2014experimental}. The meaning of the different thermal management functionalities will be explained one after another throughout this paper. By using the transformation theory, the desired thermal property distribution, e.g., thermal conductivity, will be given for specifical functionalities, e.g., thermal cloak. A relative recent review on the fundamentals, application, and outlook for the thermal metamaterial can be found in \cite{wang2020thermal}. Thermal metamaterials with the electrical current control are reported in \cite{moccia2014independent,ma2014experimental}, and their applications in electronics are reviewed in \cite{dede2018thermal}. While without using the transformation theory, gradient-based topology optimization design is also applied to find the optimal thermal property to achieve same functionalities in thermal cloak \cite{fujii2018exploring} and direct current \cite{fujii2019optimizing}. Most recently, Sha et al. \cite{sha2021robustly} used the transformation theory to first identify the optimized thermal conductive property and then leverage the topological optimization to find the best RVEs. This work was extended for anisotropic space in \cite{sha2022topology} and for omnidirectionally cloaking sensors in \cite{sha2022topology2}. Instead of using transformation theory, Xu et al. \cite{xu2022level,xu2023topology} achieved thermal cloaks in the Euclidean spaces of 2D planar surfaces and 3D solids incorporating the conformal geometry theory.

Nevertheless, several shortcomings exist in these state-of-the-art works: (1) in one scale (structural) design, multiple isotropic materials are often needed to achieve the functionalities (since there is no ability to tailor the material thermal conductivity) \cite{dede2014thermal,fujii2019topology}; (2) for thermal cloak, it is very hard to cloak the ``shield'' region itself \cite{xu2023topology}; (3) multiple functions are often difficult to be achieved in one single design \cite{shen2016thermal,fujii2020cloaking}; (4) numerous design variables exist in the multiscale design \cite{seo2020heat}; and (5) there is no interaction analysis between architecture and material.

This work is the first attempt to utilize the two-scale data-driven design framework for achieving a wide range of thermal functionalities including thermal cloak, thermal concentrator, and thermal rotator/reversal. A database is constructed which includes extensive RVE architected structures and their thermal conductive properties. Numerical homogenization is used to compute the thermal conductivity of each RVE. Steady-state heat conduction problems are solved at the structure scale with thermal conductive property originated from the lower unit cell scale. Sensitivity of each functionality w.r.t. the design variable, i.e., the \textit{property} itself, is derived. Optimized \textit{property} distribution is obtained by leveraging topological design to successfully achieve different thermal functionalities at the structure level. Next, final heterostructures are assembled by choosing unit cells/RVEs from the database to match with desired properties at each location. Numerical examples illustrate an excellent matching between the optimized properties and substituted ones, with their mean squared error as low as 9.3e-6 and coefficient of determination as high as 0.9998.

The organization of the paper is as follows. Numerical homogenization for thermal conduction and its upper level structural steady-state problem are introduced in Section 2. The data-driven \textit{property} design model is given in Section 3. Gradient information of different thermal functionalities w.r.t. the defined \textit{property} design variable is presented in Section 4. Numerical examples are shown in Section 5. Higher-scale structural assembly and the extraction quality analysis are presented in Section 6. Finally, our conclusions are set forth in Section 7.

\section{Steady-state heat conduction problem} 

At the higher structure scale, we will solve the pure steady-state heat conduction problem with the following Poisson equation without the body heat source term: 

\begin{equation} 
\rm{div}(\kappa T) = 0, 
\label{eq:Poisson} 
\end{equation} 
where $\kappa$ is the thermal conductivity. For the two-scale problem considered in this paper, effective thermal conductivity of the RVE is computed using the homogenization theory \cite{hassani1998review,hassani1998review2}. Similar to the elastic problem \cite{yvonnet2019computational, andreassen2014determine}, the homogenization equation associated with only the scalar temperature field can be written as: 

\begin{equation} 
\int_{V}{\tau_{,i} \kappa_{ij} T^{k}_{,j}}dV = \int_{V}{\tau_{,i}\kappa_{ij} T^{0(k)}_{,j}} dV 
\label{eq:H1} 
\end{equation} 
and  
\begin{equation} 
\kappa^{H}_{ij} = \frac{1}{|V|} \int_{V}{(T^{0(i)}_{,l}-T^{(i)}_{,l}}) \kappa_{lm} (T_{,m}^{0(j)} - T^{(j)}_{,m}) dV 
\label{eq:H2} 
\end{equation} 
where ${\tau}$ and $T$ are the virtual temperature and temperature field, respectively. For unstructured mesh, the integration should be conducted on each finite element, and the homogenized thermal conductivity will be written as 

\begin{equation} 
\bm{\kappa}^{H} = \frac{1}{|V|} \sum_{e=1}^{N_e}\int_{V_e}(\bm{I} - \bm{B}_{e} \bm{T}_{e} )^{T} \bm{\kappa}^{e} (\bm{I} - \bm{B}_{e} \bm{T}_{e} ) dV_{e} 
\label{eq:H3} 
\end{equation} 
where $V$ is the volume of the RVE, $N_e$ is the number of elements inside, $\bm I$ is a two times two identity matrix in 2D, and $\bm{B}_e =\bm{ L}_e \bm{N}_e$ in which $\bm{L}_e$ is the differential operator and $\bm{N}_{e}$ is the shape function matrix. $\bm{T}_{e}$ contains two columns corresponding to the two temperature fields resulting from globally enforcing the following unit temperature gradient fields: 

\begin{equation} 
\varepsilon_{T}^{1} = (1,0)^{T}, \ \ \text{and} \ \ \varepsilon_{T}^{2} = (0,1)^{T} 
\label{eq:H4} 
\end{equation} 
$\bm{T}^{0}$ in (\ref{eq:H1}) and (\ref{eq:H2}) contains the two temperature fields corresponding to the unit temperature gradient fields in (\ref{eq:H4}). The indices in parentheses in (\ref{eq:H1}) and (\ref{eq:H2}) refer to the column number.

\section{Data-driven \textit{property} design model} 

\label{Model} 

\begin{figure}[t!] 
\captionsetup{singlelinecheck = true, justification=justified} 
\centering 
\includegraphics[width=0.55\linewidth]{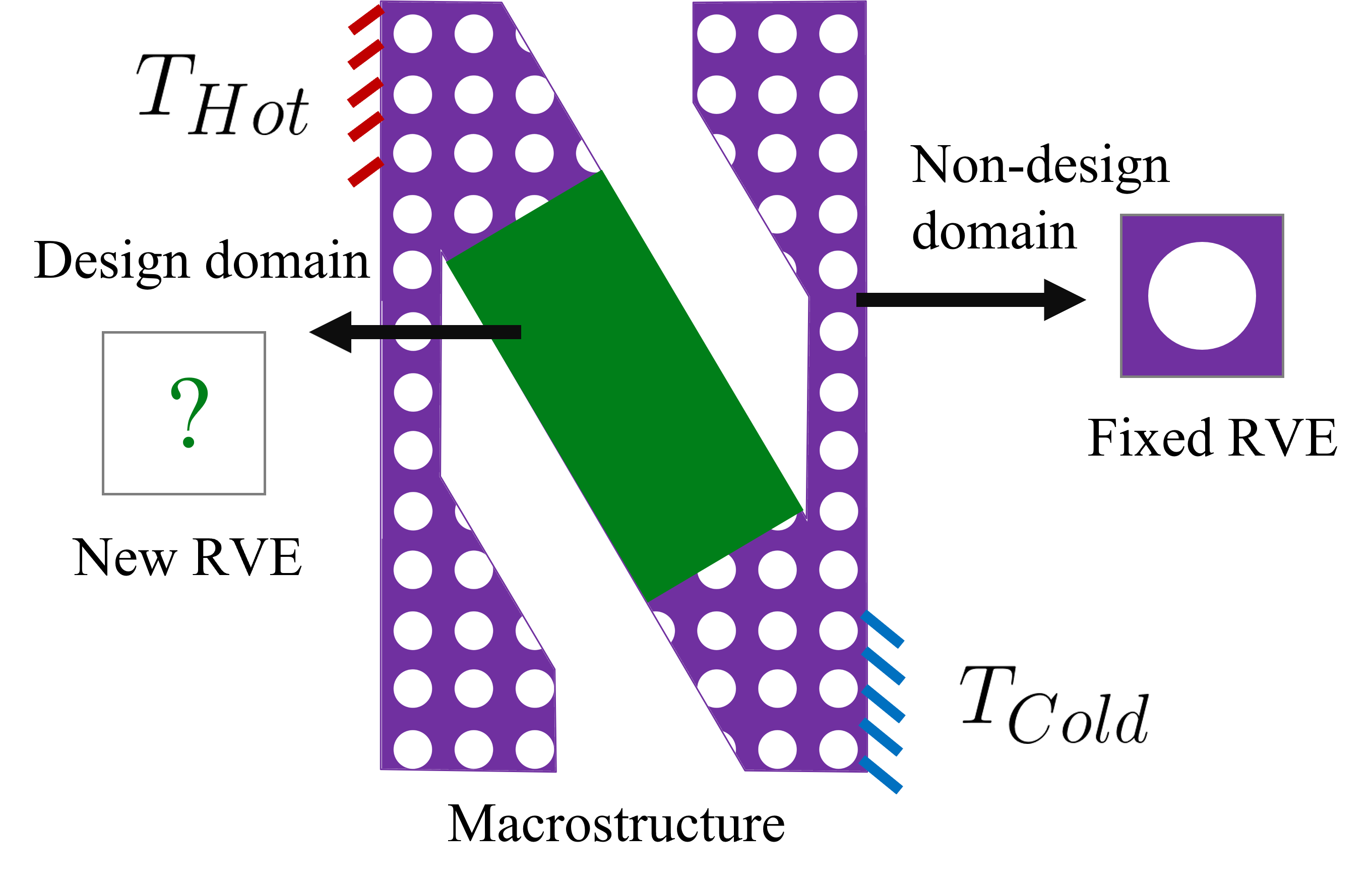} 
\caption{Schematic illustration of a macrostructure composed of design and non-design domains with various kinds of RVEs.} 
\label{fig:NU} 
\end{figure}

In the classical topology optimization model, design variables are often set as the pseudo density of each element at either the structural or material RVE level (or both in concurrent design problems). By contrast, design variables in our data-driven property design model are explicitly set as the components of the homogenized thermal conductivity tensor, as obtained using the method described in the above section. As illustrated in Fig.~\ref{fig:NU}, the topology of macrostructure in our design model is assumed to be given, while only RVEs inside the \textit{Design Domain} are optimized by tailoring its underlying homogenized thermal conductivities. Since our study focuses only on orthotropic unit cells, two independent components $\bm{\kappa}^{11}$ and $\bm{\kappa}^{22}$ can be used to fully representing the homogenized effective thermal conductivities and the RVE itself: 

\begin{equation} 
\bm{\kappa}^{H} = \begin{bmatrix} 
\bm{\kappa}^{11} & 0 \\ 
0&\bm{\kappa}^{22} 
\end{bmatrix} 
\label{eq:Orth} 
\end{equation}

\begin{figure}[t!] 
\captionsetup{singlelinecheck = true, justification=justified} 
\centering 
\includegraphics[width=0.75\linewidth]{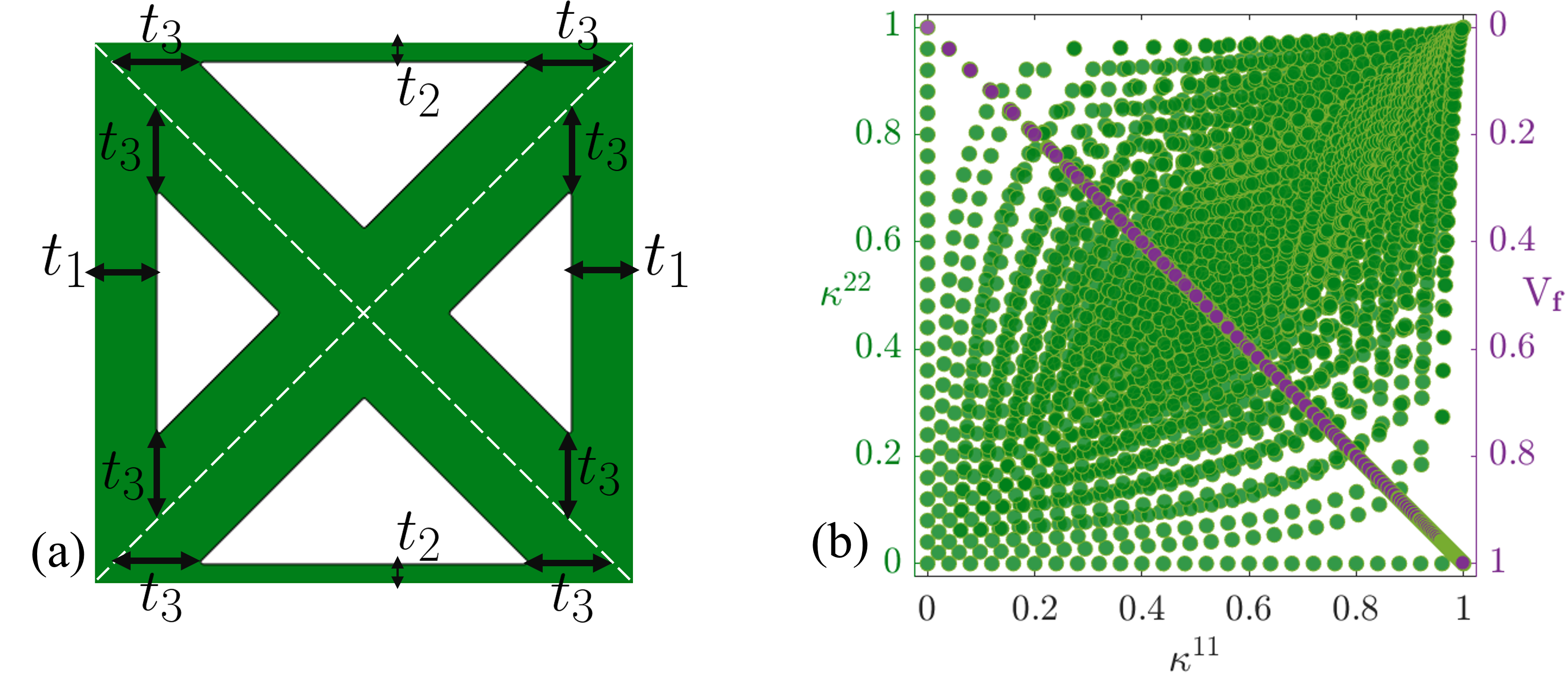} 
\caption{Left: a selected RVE and its associated three parameters ($t_{1}$, $t_2$, and $t_3$); Right: the database constructed from the left three parameters covering a wide range of the viable solutions in terms of both thermal conductive properties and volume fractions.} 
\label{fig:GeoP} 
\end{figure}

Following \cite{liu2020data}, we use three continuous variables $t_{1}$, $t_{2}$, and $t_{3}$ shown in Fig.~\ref{fig:GeoP} (a) to construct our database with different square RVEs (\textit{Geometry}) and the corresponding homogenized conductivities $\bm{\kappa}^{11}$ and $\bm{\kappa}^{22}$ (\textit{Property}). $t_{1}$ represents the width occupied by solid material on both left and right sides, $t_{2}$ represents the same for top and bottom sides, and $t_{3}$ then represents the width in the two cross or diagonal regions. Given the number of pixels or finite elements (used in numerical homogenization) in each direction, corresponding maximum values of $t_{1}$, $t_{2}$, $t_{3}$ can be identified, that is, half the number of finite elements every direction. In our database, the square RVE will be divided into $50 \times 50$ bilinear square elements, so that we have: $0 \leq t_{1} \leq 50/2$, $0 \leq t_{2} \leq 50/2$, and $0 \leq t_{3} \leq 50/2$. As a result, there are a total of $26 \times 26 \times 26 = 17576$ RVE geometries. By eliminating the repeated geometries, the total number of RVEs becomes 8282.

The thermal conduction properties as well as the volume fraction distribution of the constructed database are shown in Fig.~\ref{fig:GeoP} (b). The left side of the Y-axis in Fig.~\ref{fig:GeoP} (b) is the second component of thermal conductivity $\bm{\kappa}^{22}$, the right side is the distribution of volume fraction $V_{f}$, and the X-axis is the first component of thermal conductivity $\bm{\kappa}^{11}$. Several significant advantages are observed for our constructed database. First, the constructed database is widely distributed and covers a large portion of possible solutions. Second, because of the parameters setting, the resulting RVEs will always be well connected throughout the database. In other words, during the full structural assembling, we only need to address matching between the optimized properties and substituted ones but not the connectivity between neighboring RVEs. This is a huge advantage as our data-driven design is carried out only on properties, and there is a second step to assemble the final structure based on the properties or substituted RVEs. Finally, from the volume fraction distribution, the database covers a large number of cases from 0\% to 100\%, as can be seen from the purple points in Fig.~\ref{fig:GeoP} (b).

\begin{figure}[t!] 
\captionsetup{singlelinecheck = true, justification=justified} 
\centering 
\includegraphics[width=0.6\linewidth]{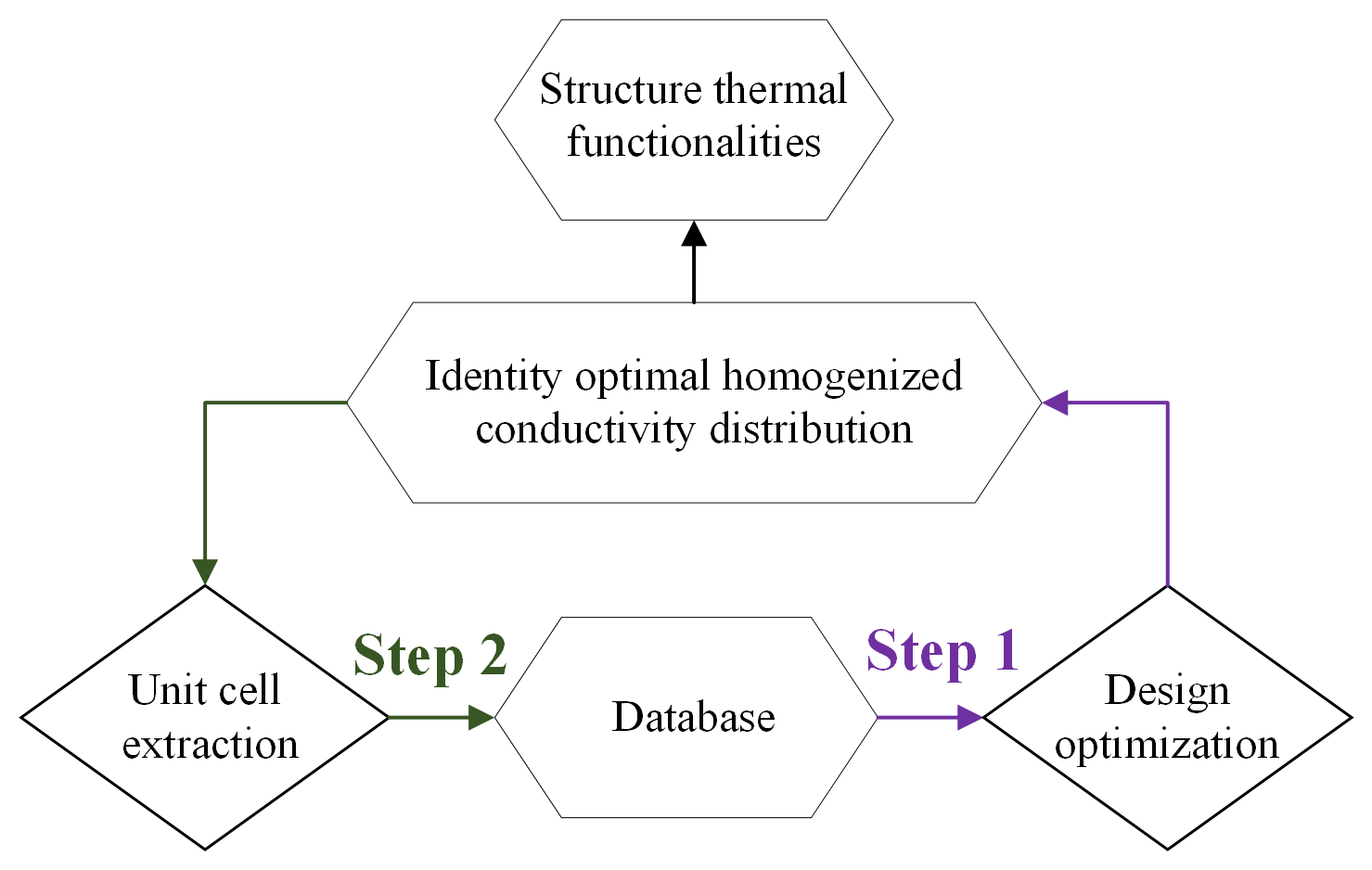} 
\caption{Two-scale data-driven \textit{property} design framework for thermal functionalities.} 
\label{Flow} 
\end{figure}

In the following sections, different objective functions associated with versatile thermal functionalities including thermal cloaks, thermal concentrators, thermal rotators, and their combinations are introduced separately. All design case studies follow the same framework as shown in Fig.~\ref{Flow}.

\subsection{Thermal cloak} 

\begin{figure}[t!] 
\captionsetup{singlelinecheck = true, justification=justified} 
\centering 
\includegraphics[width=0.96\linewidth]{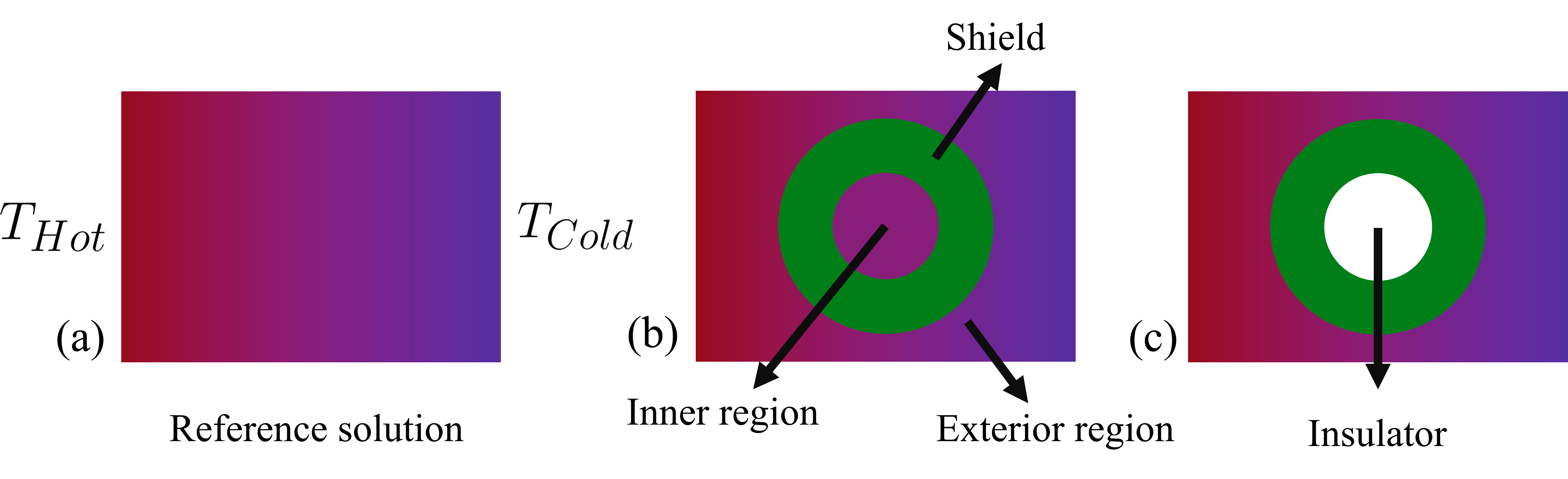} 
\caption{Schematic illustration of a thermal cloak: (a) a block of material with the reference temperature distribution; (b) a thermal cloak with non temperature gradient in the inner region and without temperature deviation in the exterior region comparing to (a); (c) a thermal cloak with a insulator in the inner region and without temperature deviation in the exterior region comparing to (a).} 
\label{fig:cloak} 
\end{figure}

A heat cloak can be formulated in a number of ways depending on its focus domain. In most of the literatures using transformation theory or non-gradient based design \cite{narayana2012heat}, a thermal cloak requires the interior hollow region introduced by the ``shield" to exhibit no temperature gradient, and the exterior region outside the ``shield" to have the same temperature as the one without the ``shield", as illustrated in Fig.~\ref{fig:cloak}. In Fig.~\ref{fig:cloak} (a), for a block of material without any ``shield", the profile has a uniform temperature gradient. This profile serves as the reference solution during our later comparisons. By contrast, in Fig.~\ref{fig:cloak} (b) in which the ``shield" is included (the green area), there is no temperature gradient in the inner hollow region, and there is no temperature change in the exterior region compared with Fig.~\ref{fig:cloak} (a).

In most gradient-based design optimizations for thermal cloak \cite{fujii2019topology,xu2023topology}, the aforementioned inner region is often replaced by an insulator as shown in Fig.~\ref{fig:cloak} (c), and the focus area can then only be the exterior region outside the ``shield". In fact, as stated in \cite{shen2016thermal}, the original intention of designing such a thermal cloak is to hide an object inside the ``shield" from the detection by only measuring the external temperature, thereby making the object thermally ``invisible". In the second formulation with the thermal insulator, when we guarantee that the external temperature is unchanged compared to the reference profile (by designing the ``shield"), we achieve the goal of cloaking and can place any objects inside the insulator to make it ``invisible". Therefore, in this study which indeed uses gradient-based numerical optimization, we take the second formulation to design the thermal cloak. While we use the data-driven property design, the mathematical formulation of the design problem would be:

\begin{align} 
\text{Minimize}: \ & J_{ck}({\boldsymbol \kappa^{11}},{\boldsymbol \kappa^{22}},{\bf T}) \label{Jck} \\ 
\boldsymbol \kappa^{11}(\mathbf x) & \\ 
\boldsymbol \kappa^{22}(\mathbf x) & \\ 
\mathbf T \in \mathcal S_T & \\ 
\text{subject}:\ & \rm{div}(\boldsymbol \kappa^{H} \ T) = 0, \label{conr1}\\ 
& T = T_{Hot}, \ \text{on} \ \Gamma_{L} \label{conr2}\\ 
\ & \ T = T_{Cold}, \ \text{on} \ \Gamma_{R}, \label{sec5:eq:opt-b}\\ 
\ & \kappa^{11}_{Min} \leq \kappa^{11} \leq \kappa^{11}_{Max},\label{sec5:pereqopt5}\\ 
\ & \kappa^{22}_{Min} \leq \kappa^{22} \leq \kappa^{22}_{Max}, \ \label{sec5:pereqopt6} 
\end{align} 
with   
\begin{equation} 
J_{ck} = \int_{\Omega_{Exterior}} \left| \frac{\bm{T} - \bm{T}_{Ref}}{ \bm{T}_{Ref}}\right|^{2} d\Omega_{Exterior}, 
\label{aa} 
\end{equation} 
where $\boldsymbol \kappa^{11}_{Min}$, $\boldsymbol \kappa^{11}_{Max}$, $\boldsymbol \kappa^{22}_{Min}$, and $\boldsymbol \kappa^{22}_{Max}$ are identified given a database containing both geometrical information and their corresponding effective thermal conductive properties. From our database, since the constitutive isotropic material is assumed with the thermal conductivity as $\kappa = 1$, the smallest and largest values for both $\kappa^{11}$ and $\kappa^{22}$ are 0 and 1, respectively. $\bm T$ is the temperature distribution of the structure with the ``shield", while $\bm{T}_{Ref}$ is the reference solution without ``shield". The objective function $J_{ck}$ would then minimize the difference of the temperature with and without the ``shield" for certain interested area, e.g., the mentioned exterior region shown in Fig.~\ref{fig:cloak} (b).

Clearly, the above objective function cannot guarantee the temperature distribution inside the ``shield" area, in which the temperature would be significantly different than the same domain from the reference solution. This issue has been identified as one of the biggest challenges in using transformation theory or non-gradient base design \cite{huang2019theoretical}, that is, the ``shield" area itself cannot be ``cloaked". Fortunately, this problem can be simply addressed within our framework by defining a different objective function as: 

\begin{equation} 
J_{ck} = \int_{\Omega_{(Exterior \ + \ Shield)}} \left| \frac{\bm{T} - \bm{T}_{Ref}}{ \bm{T}_{Ref}}\right|^{2} \ \ d\Omega_{(Exterior \ + \ Shield)} 
\label{aa-2} 
\end{equation}  
The performance of our design optimization for different objective functions or cloaked areas is shown as a numerical example in Section \ref{Ep}. To the best of our knowledge, it is also the first time gradient-based data-driven design is used to cloak the Shield/Design domain itself.


\subsection{Thermal concentrator} 

\label{TC}

\begin{figure}[t!] 
\captionsetup{singlelinecheck = true, justification=justified} 
\centering 
\includegraphics[width=0.9\linewidth]{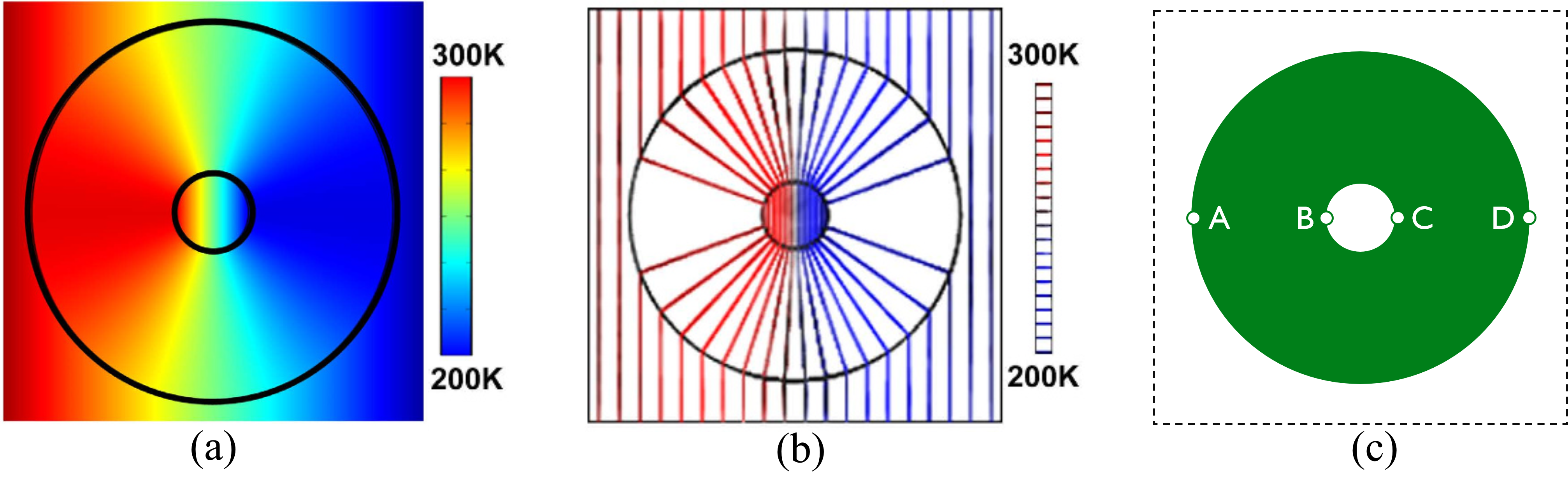} 
\caption{Illustration of a thermal concentrator from Ref. \cite{chen2015experimental}: (a) temperature profile of a thermal concentrator; (b) isothermal lines of a thermal concentrator; (c) four points at the horizontal center line to formulate the concentration efficiency.} 
\label{fig:tc} 
\end{figure}

The second interesting thermal functionality is called thermal concentration. Fig.~\ref{fig:tc} (a) from \cite{chen2015experimental} shows the temperature profile of a thermal concentrator, in which the heat is concentrated in the inner region. This phenomenon or functionality can also be observed from Fig.~\ref{fig:tc} (b) with different isothermal lines. An isothermal line is a line along which the temperature is the same. Following again the Ref.~\cite{chen2015experimental}, a thermal concentration index based on the temperature values of four different points A, B, C, and D in Fig.~\ref{fig:tc} (c) is established:

\begin{equation} 
J_{ct} = \left | \frac{T_{B} - T_{C}}{T_{A} - T_{D}} \right | 
\label{tct} 
\end{equation} 
For an ideal metamaterial device with thermal concentration, the above $J_{ct}$ should be equal to 100\%. The $J_{ct}$ value for the temperature profile in Figs.~\ref{fig:tc} (a) and (b) is 96.3\%.  

In our data-driven \textit{property} design model for the thermal concentrator, the mathematical problem can then be formulated as: 

\begin{align} 
\text{Minimize}: \ & \left( J_{ct}({\boldsymbol \kappa^{11}},{\boldsymbol \kappa^{22}},{\bf T})-100\% \right) ^2 \nonumber \\ 
\boldsymbol \kappa^{11}(\mathbf x) & \nonumber \\ 
\boldsymbol \kappa^{22}(\mathbf x) & \nonumber \\ 
\mathbf T \in \mathcal S_T & \nonumber \\ 
\text{subject}:\ & \rm{div}(\boldsymbol \kappa^{H} \ T) = 0, \\ 
& T = T_{Hot}, \ \text{on} \ \Gamma_{L},\nonumber \\ 
\ & \ T = T_{Cold}, \ \text{on} \ \Gamma_{R}, \label{sec5:eq:opt-b} \nonumber \\ 
\ & \kappa^{11}_{Min} \leq \kappa^{11} \leq \kappa^{11}_{Max}, \nonumber \\ 
\ & \kappa^{22}_{Min} \leq \kappa^{22} \leq \kappa^{22}_{Max} \nonumber \  
\end{align}

\subsection{ Thermal rotator/inverter} 

\label{TRI}

The third thermal functionality is called thermal rotator or even thermal inverter in the limit state. Before introducing this functionality, a fundamental functional named thermal or heat flux should be given:

\begin{equation} 
\overrightarrow{ \bm {\phi}_{q}} = - \bm{\kappa} \nabla \bm{T} 
\label{hf} 
\end{equation} 
The heat flux is a vector quantity, and the negative sign indicates that heat flux moves from the hot temperature region to the low temperature region.

\begin{figure}[t!] 
\captionsetup{singlelinecheck = true, justification=justified} 
\centering 
\includegraphics[width=0.8\linewidth]{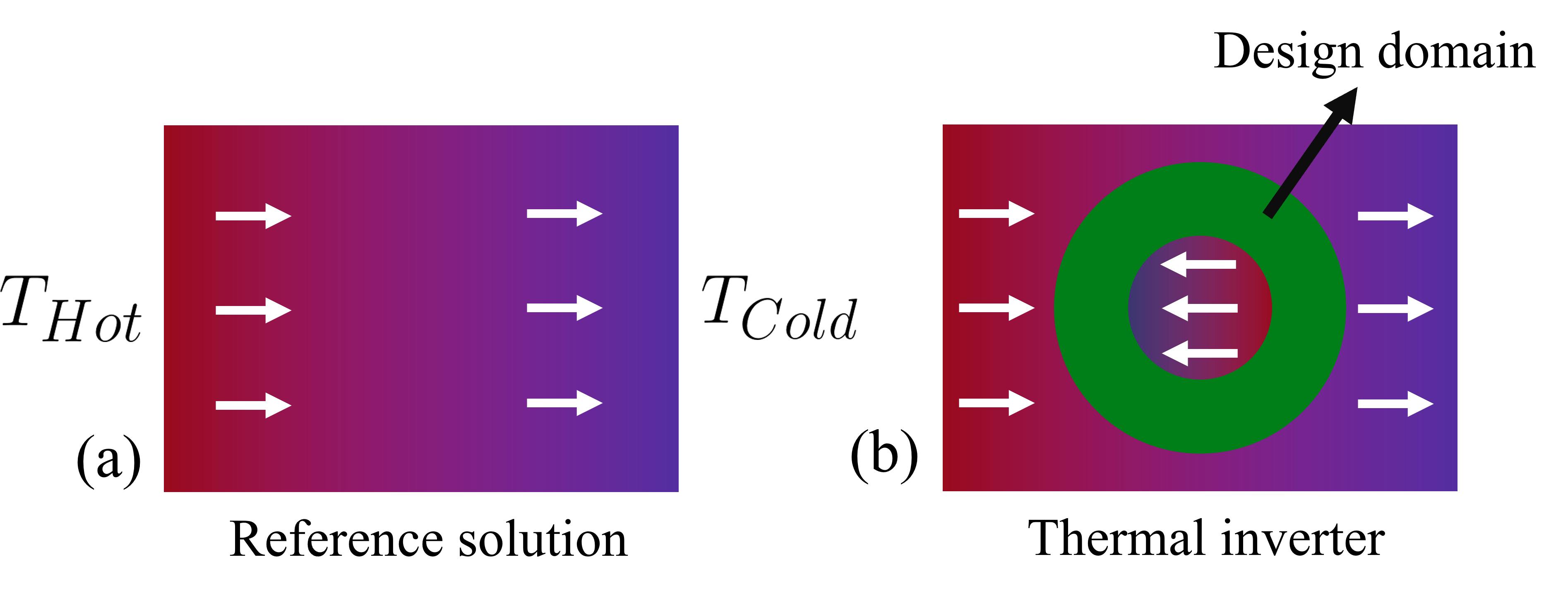} 
\caption{Schematic Illustration of a thermal rotator/inverter: a block of material with the reference temperature profile; (b) a thermal rotator/inverter.} 
\label{fig:ti} 
\end{figure}

As shown in Fig.~\ref{fig:ti} (a), when the hot source is on the left boundary and the cold source is on the right, the heat flux will move from left to right, i.e., from higher regions to lower regions throughout the whole domain. The goal of a thermal rotator/inverter is to manipulate the temperature distribution so that in certain regions, the heat flux will move from right to left, without changing the location of the hot and cold sources. This is a metamaterial property because if there is no design, the left region which is closer to the hot source will always have a higher temperature (than the right region), and the heat flux throughout the domain will always move from left to right.

To achieve the above goal, we select the objective function associated with a unit direction vector from left to right $\bm {q_{e}} $ and the heat flux of the interested area $\Omega_{obj}$ in which the heat flux is to be rotated or inverted:

\begin{equation} 
J_{ri} = \sum_{\omega \in \Omega_{obj}} {\hat {\bm {q}} \ \cdot \ \overrightarrow{ \bm {\phi}_{q}^{\omega}}}  
\label{tri} 
\end{equation} 
The dot production between the interested heat flux $\overrightarrow{ \bm {\phi}_{q}^{\omega}}$ and unit vector $\hat {\bm {q}} = [1 \ \ 0]$ will result a positive value if $\overrightarrow{ \bm {\phi}_{q}^{\omega}}$ has the exact same direction as $\hat {\bm {q}}$, i.e., from left to right. By minimizing this objective function, we are able to invert the interested heat flux to achieve a negative objective function value. Hence, our data-driven \textit{property} design problem can be mathematically formulated as: 

\begin{align} 
\text{Minimize}: \ & J_{ri}({\boldsymbol \kappa^{11}},{\boldsymbol \kappa^{22}},{\bf T}) \nonumber \\ 
\boldsymbol \kappa^{11}(\mathbf x) & \nonumber \\ 
\boldsymbol \kappa^{22}(\mathbf x) & \nonumber \\ 
\mathbf T \in \mathcal S_T & \nonumber \\ 
\text{subject}:\ & \rm{div}(\boldsymbol \kappa^{H} \ T) = 0, \\ 
& T = T_{Hot}, \ \text{on} \ \Gamma_{L},\nonumber \\ 
\ & \ T = T_{Cold}, \ \text{on} \ \Gamma_{R}, \label{sec5:eq:opt-b}\nonumber \\ 
\ & \kappa^{11}_{Min} \leq \kappa^{11} \leq \kappa^{11}_{Max},\nonumber \\ 
\ & \kappa^{22}_{Min} \leq \kappa^{22} \leq \kappa^{22}_{Max} \nonumber \  
\end{align}


\subsection{ Multiple thermal functionalities} 

\label{MF}

One of the biggest advantages of using our data-driven gradient-based design is that we can pursue multiple (meta-) thermal functionalities simply by assigning multiple objective functions. In this case, the design optimization problem can be mathematically formulated as 

\begin{align} 
\text{Minimize}: \ & \xi_{ck}\frac{J_{ck}}{J_{ck}^{0}} + \xi_{ct} \frac{(J_{ct} -100\%)^2}{(J_{ct}^{0} -100\%)^2}+ \xi_{ri} \frac{J_{ri}}{J_{ri}^{0}} \nonumber \\ 
\boldsymbol \kappa^{11}(\mathbf x) & \nonumber \\ 
\boldsymbol \kappa^{22}(\mathbf x) & \nonumber \\ 
\mathbf T \in \mathcal S_T & \nonumber \\ 
\text{subject}:\ & \rm{div}(\boldsymbol \kappa^{H} \ T) = 0, \label{MTF} \\ 
& T = T_{Hot}, \ \text{on} \ \Gamma_{L},\nonumber \\ 
\ & \ T = T_{Cold}, \ \text{on} \ \Gamma_{R}, \label{sec5:eq:opt-b}\nonumber \\ 
\ & \kappa^{11}_{Min} \leq \kappa^{11} \leq \kappa^{11}_{Max},\nonumber \\ 
\ & \kappa^{22}_{Min} \leq \kappa^{22} \leq \kappa^{22}_{Max} \nonumber \  
\end{align} 
where $J_{ck}^{0}$, $J_{ct}^{0}$, and $J_{ri}^{0}$ are the normalization terms and the corresponding index values for thermal cloak, concentrator, and rotator before the optimization. $\xi_{ck}$, $\xi_{ct}$, and $\xi_{ri}$ are different weights for different thermal functionalities used in the optimization with multiple objective functions.


\section{Sensitivity Analysis}  

\label{SA}

To conduct the proposed gradient-based data-driven \textit{property} design, the gradient information of the selected objective function w.r.t. the design variables (components of the thermal conductivity $ \bm \kappa^{11}$ and $\bm \kappa^{22}$) should be derived. This gradient is often called sensitivity. To illustrate the sensitivity analysis, we use the discrete equilibrium equation for the steady state thermal conduction problem Eq. (\ref{eq:Poisson}): 

\begin{equation} 
{ {\bm K}_{T}} { \bm {T}} = { \bm {F}_{T}} 
\label{Tee} 
\end{equation} 
where $ {\bm K}_{T}$ is the global thermal stiffness matrix, $\bm {T}$ is the unknown temperature profile, and ${\bm F}_{T}$ represents the global temperature load vector. The global thermal stiffness matrix $ {\bm K}_{T}$ can be assembled from elemental thermal stiffness matrix $ {\bm K}_{e}$ in a standard finite element way. For each finite element, $ {\bm K}_{e}$ can be assembled by: 

\begin{equation} 
{\bm K}_{e} = \int_{\Omega_{e}} \bm{B}_{e} \bm{\kappa}^{H}_{e} \bm{B}_{e} d\Omega_{e} 
\label{eq:KTe} 
\end{equation} 
where $\bm{\kappa}^{H}_{e}$ is computed from Eq. (\ref{eq:H3}). In the following, the sensitivity of different/multiple functionalities index w.r.t. the \textit{property} design variable is derived.


\subsection{Thermal cloak}

For thermal cloak, the objective function (\ref{Jck}) can be written as a Lagrangian augmented function by adding the thermal equilibrium equation (\ref{Tee}): 

\begin{equation} 
J^{\mathcal{L}}_{ck} = J_{ck}({\boldsymbol \kappa^{11}},{\boldsymbol \kappa^{22}},{\bf T}) - \bm{ \lambda}^{T} \left( { {\bm K}_{T} ({\boldsymbol \kappa^{11}},{\boldsymbol \kappa^{22}}) } \ { \bm {T}} - { \bm {F}_{T}} \right)  
\end{equation} 
where $\bm{ \lambda}^{T} $ is the so-called Lagrange multiplier as a vector. Taking the objective function Eq.~(\ref{aa}) as an example, the above equation can be further written as:

\begin{equation} 
J^{\mathcal{L}}_{ck} = \left(\frac{\bm{T} - \bm{T}_{Ref}}{ \bm{T}_{Ref}}\right)^{T} {\bm D} \ (\frac{\bm{T} - \bm{T}_{Ref}}{ \bm{T}_{Ref}}) - \bm{ \lambda}^{T} \left( { {\bm K}_{T} ({\boldsymbol \kappa^{11}},{\boldsymbol \kappa^{22}}) } \ { \bm {T}} - { \bm {F}_{T}} \right)  
\end{equation} 
where ${\bm D} $ a diagonal matrix associated with the query nodes of the interested cloak domain. Taking the partial derivative of the Lagrangian $J^{\mathcal{L}}_{ck}$ w.r.t. the thermal conduction component ${ \kappa^{11}_{e}}$ as an example, we have: 

\begin{align} 
\frac{\partial J^{\mathcal{L}}_{ck}}{ \partial \kappa^{11}_{e}} & = 2 {\bm D} \left(\frac{\bm{T} - \bm{T}_{Ref}}{ \bm{T}_{Ref}^2} \right) \frac{\partial \bm{T}}{\partial \kappa^{11}_{e}} - \bm{ \lambda}^{T} \left( \frac{\partial{{\bm K}_{T}}} {\partial { \kappa^{11}_{e}}} \bm{T} + {\bm K}_{T} \frac{\partial \bm T} {\partial { \kappa^{11}_{e}}} \right) \\ 
\ & = 2 {\bm D} \left(\frac{\bm{T} - \bm{T}_{Ref}}{ \bm{T}_{Ref}^2}\right) \frac{\partial \bm{T}}{\partial \kappa^{11}_{e}} - \bm{ \lambda}^{T} \frac{\partial{{\bm K}_{T}}} {\partial { \kappa^{11}_{e}}} \bm{T} - \bm{ \lambda}^{T} {\bm K}_{T} \frac{\partial \bm T} {\partial { \kappa^{11}_{e}}} \\ 
\ & = \left( 2 {\bm D} \left(\frac{\bm{T} - \bm{T}_{Ref}}{ \bm{T}_{Ref}^2}\right) - \bm{ \lambda}^{T} {\bm K}_{T} \right) \frac{\partial \bm{T}}{\partial \kappa^{11}_{e}} - \bm{ \lambda}^{T} \frac{\partial{{\bm K}_{T}}} {\partial { \kappa^{11}_{e}}} \bm{T} 
\label{29} 
\end{align} 
The first term on the right-hand side of (\ref{29}) can be eliminated by setting the following

\begin{equation} 
2 {\bm D} \left(\frac{\bm{T} - \bm{T}_{Ref}}{ \bm{T}_{Ref}^2}\right) - \bm{ \lambda}^{T} {\bm K}_{T} = 0 
\label{eq:adj} 
\end{equation} 
Through solving the adjoint equation (\ref{eq:adj}), we can get the multiplier as 

\begin{equation} 
\bm{ \lambda}^{T} = 2 {\bm D} \left(\frac{\bm{T} - \bm{T}_{Ref}}{ \bm{T}_{Ref}^2}\right) {\bm K}_{T}^{-1} 
\label{eq:adj2} 
\end{equation} 
Finally, the derived sensitivity can be written as

\begin{equation} 
\frac{\partial J^{\mathcal{L}}_{ck}}{ \partial \boldsymbol \kappa^{11}} = 2 {\bm D} \left(\frac{\bm{T} - \bm{T}_{Ref}}{ \bm{T}_{Ref}^2}\right) {\bm K}_{T}^{-1} \frac{\partial{{\bm K}_{T}}} {\partial { \kappa^{11}_{e}}} \bm{T} 
\label{eq:st} 
\end{equation} 
in which $\frac{\partial{{\bm K}_{T}}} {\partial { \kappa^{11}_{e}}}$ can be directly obtained through the finite element scheme operation, and vice versa for $\frac{\partial{{\bm K}_{T}}} {\partial { \kappa^{22}_{e}}}$. For different objective functions in Eqs.~(\ref{aa}) and (\ref{aa-2}), we need only to update the equation by changing the diagonal matrix ${\bm D} $.


\subsection{Thermal concentrator}

Firstly, the thermal concentrator index Eq.~(\ref{tct}) can be rewritten as: 

\begin{equation} 
J_{ct} = \left | \frac{T_{B} - T_{C}}{T_{A} - T_{D}} \right | = \left | { \frac{\bm V_{B}{\bm T} - \bm{V}_{C}{\bm T}}{\bm{V}_{A}{\bm T}- \bm{V}_{D}{\bm T}} } \right |  
\label{tct2} 
\end{equation} 
where ${\bm V}_{A}$, ${\bm V}_{B}$, ${\bm V}_{C}$, and ${\bm V}_{D}$ are row vectors associated with finite element nodes in points $A$, ${B}$, ${C}$, and ${D}$, respectively (similar to the diagonal matrix ${\bm D}$ above). Similar also to the sensitivity derivation above for the thermal cloak, we first have the Lagrangian augmented function for the thermal concentrator as 

\begin{align} 
J^{\mathcal{L}}_{ct} = &\left( J_{ct}({\boldsymbol \kappa^{11}},{\boldsymbol \kappa^{22}},{\bf T})-100\% \right)^2 - \bm{ \lambda}^{T}_{A} \left( { {\bm K}_{T} ({\boldsymbol \kappa^{11}},{\boldsymbol \kappa^{22}}) } \ { \bm {T}} - { \bm {F}_{T}} \right) \nonumber \\ 
&- \bm{ \lambda}^{T}_{B} \left( { {\bm K}_{T} ({\boldsymbol \kappa^{11}},{\boldsymbol \kappa^{22}}) } \ { \bm {T}} - { \bm {F}_{T}} \right) - - \bm{ \lambda}^{T}_{C} \left( { {\bm K}_{T} ({\boldsymbol \kappa^{11}},{\boldsymbol \kappa^{22}}) } \ { \bm {T}} - { \bm {F}_{T}} \right) \nonumber \\ 
&- \bm{ \lambda}^{T}_{D} \left( { {\bm K}_{T} ({\boldsymbol \kappa^{11}},{\boldsymbol \kappa^{22}}) } \ { \bm {T}} - { \bm {F}_{T}} \right) 
\label{ctL} 
\end{align} 
By plugging Eq.~(\ref{tct2}) in (\ref{ctL}), we have 

\begin{align} 
J^{\mathcal{L}}_{ct} = & \left( { \frac{\bm V_{B}{\bm T} - \bm{V}_{C}{\bm T}}{\bm{V}_{A}{\bm T}- \bm{V}_{D}{\bm T}} } - 1 \right)^2 - \bm{ \lambda}^{T}_{A} \left( { {\bm K}_{T} } \ { \bm {T}} - { \bm {F}_{T}} \right) - \bm{ \lambda}^{T}_{B} \left( { {\bm K}_{T} } \ { \bm {T}} - { \bm {F}_{T}} \right) \nonumber \\ 
& - \bm{ \lambda}^{T}_{C} \left( { {\bm K}_{T}} \ { \bm {T}} - { \bm {F}_{T}} \right) - \bm{ \lambda}^{T}_{D} \left( { {\bm K}_{T} } \ { \bm {T}} - { \bm {F}_{T}} \right)  
\label{ctL2} 
\end{align} 
Taking again the sensitivity of $J^{\mathcal{L}}_{ct}$ w.r.t. ${ \kappa^{11}_{e}}$ as an example, we have 

\begin{align} 
\frac{\partial J^{\mathcal{L}}_{ct}}{ \partial \kappa^{11}_{e}}  
& = \frac{ \partial \left( \frac{\bm V_{B}{\bm T} - \bm{V}_{C}{\bm T}}{\bm{V}_{A}{\bm T}- \bm{V}_{D}{\bm T}} - 1 \right)^2}{\partial \kappa^{11}_{e}} \underbrace{ - \bm{ \lambda}^{T}_{A} \left( \frac{\partial{{\bm K}_{T}}} {\partial { \kappa^{11}_{e}}} \bm{T} + {\bm K}_{T} \frac{\partial \bm T} {\partial { \kappa^{11}_{e}}} \right)}_{\Lambda_{A}} \underbrace{ - \bm{ \lambda}^{T}_{B} \left( \frac{\partial{{\bm K}_{T}}} {\partial { \kappa^{11}_{e}}} \bm{T} + {\bm K}_{T} \frac{\partial \bm T} {\partial { \kappa^{11}_{e}}} \right)}_{\Lambda_{B}} \nonumber \\ 
\ & \ \ \ \ \ \ \underbrace{ - \bm{ \lambda}^{T}_{C} \left( \frac{\partial{{\bm K}_{T}}} {\partial { \kappa^{11}_{e}}} \bm{T} + {\bm K}_{T} \frac{\partial \bm T} {\partial { \kappa^{11}_{e}}} \right)}_{\Lambda_{C}} \underbrace{ - \bm{ \lambda}^{T}_{D} \left( \frac{\partial{{\bm K}_{T}}} {\partial { \kappa^{11}_{e}}} \bm{T} + {\bm K}_{T} \frac{\partial \bm T} {\partial { \kappa^{11}_{e}}} \right)}_{\Lambda_{D}} \\  
\ & = 2 \left( \frac{\bm V_{B}{\bm T} - \bm{V}_{C}{\bm T}}{\bm{V}_{A}{\bm T}- \bm{V}_{D}{\bm T}} - 1 \right) \frac{ \partial \left( \frac{\bm V_{B}{\bm T} - \bm{V}_{C}{\bm T}}{\bm{V}_{A}{\bm T}- \bm{V}_{D}{\bm T}} \right)}{\partial \kappa^{11}_{e}} + {\Lambda_{A}} + {\Lambda_{B}} + {\Lambda_{C}} +{\Lambda_{D}} \\ 
\ & = \underbrace{2 \left( \frac{\bm V_{B}{\bm T} - \bm{V}_{C}{\bm T}}{\bm{V}_{A}{\bm T}- \bm{V}_{D}{\bm T}} - 1 \right)}_{\complement} \frac{1}{({\bm{V}_{A}{\bm T}- \bm{V}_{D}{\bm T}})^2} \left[ \left( {\bm V}_{B}\frac{\partial \bm T} {\partial { \kappa^{11}_{e}}} - {\bm V}_{C}\frac{\partial \bm T} {\partial { \kappa^{11}_{e}}} \right ) \left( {\bm{V}_{A}{\bm T}- \bm{V}_{D}{\bm T}} \right) \right. \nonumber \\ 
\ & \ \ \ \ \ \ \left. - \left( {\bm V}_{A}\frac{\partial \bm T} {\partial { \kappa^{11}_{e}}} - {\bm V}_{D}\frac{\partial \bm T} {\partial { \kappa^{11}_{e}}} \right ) \left( {\bm{V}_{B}{\bm T}- \bm{V}_{C}{\bm T}} \right) \right] + {\Lambda_{A}} + {\Lambda_{B}} + {\Lambda_{C}} +{\Lambda_{D}} \\ 
\ & = \frac{{\complement} {\bm V}_{B}}{\left( \bm V_{A}{\bm T} - \bm{V}_{D}{\bm T} \right) } \ \frac{\partial \bm T} {\partial { \kappa^{11}_{e}}} - \frac{{\complement} {\bm V}_{C}}{\left( \bm V_{A}{\bm T} - \bm{V}_{D}{\bm T} \right) } \ \frac{\partial \bm T} {\partial { \kappa^{11}_{e}}} - \frac{{\complement} {\bm V}_{A}\ ({\bm V_{B}{\bm T} - \bm{V}_{C}{\bm T}})} {\left( \bm V_{A}{\bm T} - \bm{V}_{D}{\bm T} \right)^2 } \ \frac{\partial \bm T} {\partial { \kappa^{11}_{e}}} \nonumber \\ 
\ & \ \ \ \ \ \ + \frac{{\complement} {\bm V}_{D}\ ({\bm V_{B}{\bm T} - \bm{V}_{C}{\bm T}})} {\left( \bm V_{A}{\bm T} - \bm{V}_{D}{\bm T} \right)^2 } \ \frac{\partial \bm T} {\partial { \kappa^{11}_{e}}} + {\Lambda_{A}} + {\Lambda_{B}} + {\Lambda_{C}} +{\Lambda_{D}} \\ 
\ & = \underbrace{ \left[ \frac{{\complement} {\bm V}_{B}}{\left( \bm V_{A}{\bm T} - \bm{V}_{D}{\bm T} \right) } - \bm{ \lambda}^{T}_{A} {\bm K}_{T} \right]}_{\mathcal{A}_{B}} \ \frac{\partial \bm T} {\partial {\boldsymbol \kappa^{11}}} - \bm{ \lambda}^{T}_{B} \frac{\partial{{\bm K}_{T}}} {\partial { \kappa^{11}_{e}}} \bm{T} \nonumber \\ 
\ & \ \ \ \ \ \ + \underbrace{\left[ \frac{-{\complement} {\bm V}_{C}}{\left( \bm V_{A}{\bm T} - \bm{V}_{D}{\bm T} \right) } - \bm{ \lambda}^{T}_{C} {\bm K}_{T} \right]}_{\mathcal{A}_{C}} \ \frac{\partial \bm T} {\partial { \kappa^{11}_{e}}} - \bm{ \lambda}^{T}_{C} \frac{\partial{{\bm K}_{T}}} {\partial { \kappa^{11}_{e}}} \bm{T} \nonumber \\  
\ & \ \ \ \ \ \ \ \ \ \ + \underbrace{\left[ \frac{-{\complement} {\bm V}_{A} ({\bm V_{B}{\bm T} - \bm{V}_{C}{\bm T}}) } {\left( \bm V_{A}{\bm T} - \bm{V}_{D}{\bm T} \right)^2 } - \bm{ \lambda}^{T}_{A} {\bm K}_{T} \right]}_{\mathcal{A}_{A}} \ \frac{\partial \bm T} {\partial { \kappa^{11}_{e}}} - \bm{ \lambda}^{T}_{A} \frac{\partial{{\bm K}_{T}}} {\partial { \kappa^{11}_{e}}} \bm{T} \nonumber \\ 
\ & \ \ \ \ \ \ \ \ \ \ \ \ \ \ + \underbrace{\left[ \frac{{\complement} {\bm V}_{D} ({\bm V_{B}{\bm T} - \bm{V}_{C}{\bm T}}) } {\left( \bm V_{A}{\bm T} - \bm{V}_{D}{\bm T} \right)^2 } - \bm{ \lambda}^{T}_{D} {\bm K}_{T} \right]}_{\mathcal{A}_{D}} \ \frac{\partial \bm T} {\partial { \kappa^{11}_{e}}} - \bm{ \lambda}^{T}_{D} \frac{\partial{{\bm K}_{T}}} {\partial { \kappa^{11}_{e}}} \bm{T} \label{sct} 
\end{align} 
By solving the adjoint equations ${\mathcal{A}_{A}}$, ${\mathcal{A}_{B}}$, ${\mathcal{A}_{C}}$, and ${\mathcal{A}_{D}}$, we can eliminate the unknown term $\frac{\partial \bm T} {\partial { \kappa^{11}_{e}}}$ and obtain the Lagrange multipliers $ \bm{ \lambda}^{T}_{A} $, $ \bm{ \lambda}^{T}_{B} $, $ \bm{ \lambda}^{T}_{C} $, and $ \bm{ \lambda}^{T}_{D} $: 

\begin{align} 
\text{let} \ {\mathcal{A}_{A}} & = 0, \ \text{with} \ \bm{ \lambda}^{T}_{A} = \frac{-{\complement} {\bm V}_{A} ({\bm V_{B}{\bm T} - \bm{V}_{C}{\bm T}}) } {\left( \bm V_{A}{\bm T} - \bm{V}_{D}{\bm T} \right)^2 } {\bm K}_{T}^{-1}, \\ 
\text{let} \ {\mathcal{A}_{B}} & = 0, \ \text{with} \ \bm{ \lambda}^{T}_{B} = \frac{{\complement} {\bm V}_{B}}{\left( \bm V_{A}{\bm T} - \bm{V}_{D}{\bm T} \right) } {\bm K}_{T} ^{-1} , \\ 
\text{let} \ {\mathcal{A}_{C}} & = 0, \ \text{with} \ \bm{ \lambda}^{T}_{C} = \frac{-{\complement} {\bm V}_{C}}{\left( \bm V_{A}{\bm T} - \bm{V}_{D}{\bm T} \right) } {\bm K}_{T} ^{-1} , \\ 
\text{let} \ {\mathcal{A}_{D}} & = 0, \ \text{with} \ \bm{ \lambda}^{T}_{D} = \frac{{\complement} {\bm V}_{D} ({\bm V_{B}{\bm T} - \bm{V}_{C}{\bm T}}) } {\left( \bm V_{D}{\bm T} - \bm{V}_{D}{\bm T} \right)^2 } {\bm K}_{T}^{-1}. 
\end{align} 
Finally, the sensitivity Eq.~(\ref{sct}) can be written as: 

\begin{align} 
\frac{\partial J^{\mathcal{L}}_{ct}}{ \partial \kappa^{11}_{e}} = {2 \left( \frac{\bm V_{B}{\bm T} - \bm{V}_{C}{\bm T}}{\bm{V}_{A}{\bm T}- \bm{V}_{D}{\bm T}} - 1 \right)} \left[ \left( \frac{{\bm V}_{A} ({\bm V_{B}{\bm T} - \bm{V}_{C}{\bm T}}) } {\left( \bm V_{A}{\bm T} - \bm{V}_{D}{\bm T} \right)^2 } \right) {\bm K}_{T}^{-1} -  
\left( \frac{ {\bm V}_{B}}{\left( \bm V_{A}{\bm T} - \bm{V}_{D}{\bm T} \right) } \right) {\bm K}_{T} ^{-1} \right. \nonumber \\  
\left. + \left( \frac{{\bm V}_{C}}{\left( \bm V_{A}{\bm T} - \bm{V}_{D}{\bm T} \right) } \right) {\bm K}_{T} ^{-1} - \left( \frac{ {\bm V}_{D} ({\bm V_{B}{\bm T} - \bm{V}_{C}{\bm T}}) } {\left( \bm V_{A}{\bm T} - \bm{V}_{D}{\bm T} \right)^2 } \right) {\bm K}_{T}^{-1} \right] \frac{\partial{{\bm K}_{T}}} {\partial { \kappa^{11}_{e}}} \bm{T} 
\label{tctf} 
\end{align} 
Correspondingly, the sensitivity $\frac{\partial J^{\mathcal{L}}_{ct}}{ \partial \kappa^{22}_{e}}$ can be obtained through (\ref{tctf}) by only replacing $\frac{\partial{{\bm K}_{T}}} {\partial { \kappa^{11}_{e}}}$ with $\frac{\partial{{\bm K}_{T}}} {\partial { \kappa^{22}_{e}}}$.


\subsection{Thermal rotator/inverter} 

For heat flux rotation and inverse, the temperature gradient $\nabla \bm{T}$ in the heat flux computation can be written as: 

\begin{align} 
\begin{cases} \overrightarrow{ \bm {\phi}_{q}} = - \bm{\kappa} \nabla \bm{T} \\[4pt] 
\nabla \bm{T} = \begin{Bmatrix}{\frac{\partial \bm{T}}{\partial x}}\\[4pt] {\frac{\partial \bm{T}}{\partial y}}\end{Bmatrix} = \begin{Bmatrix}{\frac{\partial \bm{N}}{\partial x}} \\[4pt] {\frac{\partial \bm{N}}{\partial y}} \end{Bmatrix} {\bm T}^{(e)} = {\bm B} {\bm T}^{(e)}, \ \ \ \  
{\bm B} = \begin{bmatrix} \frac{\partial{N_1}}{\partial x} & \frac{\partial{N_2}}{\partial x} & \frac{\partial{N_3}}{\partial x} & \frac{\partial{N_4}}{\partial x}\\[4pt] \frac{\partial{N_1}}{\partial y} & \frac{\partial{N_2}}{\partial y} & \frac{\partial{N_3}}{\partial y} & \frac{\partial{N_4}}{\partial y}\\ \end{bmatrix}  
\end{cases} 
\end{align} 
$\bm{N}$ is the shape function of the adopted four-node bilinear square element. The Lagrangian for the thermal rotator/inverter is: 

\begin{align} 
J^{\mathcal{L}}_{ri} = \sum_{\omega \in \Omega_{obj}} {\hat {\bm {q}} \ \cdot \ \overrightarrow{ \bm {\phi}_{q}^{\omega}}} - \ \bm{ \lambda}^{T} \left( { {\bm K}_{T} } \ { \bm {T}} - { \bm {F}_{T}} \right) 
\end{align} 
Then, we have

\begin{align} 
\frac{\partial J^{\mathcal{L}}_{ri}}{ \partial \kappa^{11}_{e}} = & \left [ \sum_{\omega \in \Omega_{obj}} \left( \frac{\partial \hat {\bm {q}}}{\partial \kappa^{11}_{e}} \ \cdot \ \overrightarrow{ \bm {\phi}_{q}^{\omega}} + \hat {\bm {q}} \ \cdot \ \frac{\partial \overrightarrow{ \bm {\phi}_{q}^{\omega}}}{\partial \kappa^{11}_{e}} \right) \right] - 
\bm{ \lambda}^{T} \underbrace{\left( \frac{\partial{{\bm K}_{T}}} {\partial { \kappa^{11}_{e}}} \bm{T} + {\bm K}_{T} \frac{\partial \bm T} {\partial { \kappa^{11}_{e}}} \right)}_{0} 
\end{align} 
If we let the second term on the RHS above equal to 0, we have

\begin{align} 
\frac{\partial \bm T} {\partial { \kappa^{11}_{e}}} \ = & - {\bm K}_{T}^{-1}\frac{\partial{{\bm K}_{T}}} {\partial { \kappa^{11}_{e}}} \bm{T} \ \ \ \ \text{and} \\ 
\frac{\partial J^{\mathcal{L}}_{ri}}{ \partial \kappa^{11}_{e}} = & \ \sum_{\omega \in \Omega_{obj}} \left( \cancelto{0}{\frac{\partial \hat {\bm {q}}}{\partial \kappa^{11}_{e}} \ \cdot \ \overrightarrow{ \bm {\phi}_{q}^{\omega}}} + \hat {\bm {q}} \ \cdot \ \frac{\partial \overrightarrow{ \bm {\phi}_{q}^{\omega}}}{\partial \kappa^{11}_{e}} \right) \nonumber \\ 
\ = & \ \sum_{\omega \in \Omega_{obj}} \left( - \hat {\bm {q}} \ \cdot \ \frac{\partial \bm{\kappa}^{(\omega)}}{\partial \kappa^{11}_{e}} \nabla \bm{T}^{(\omega)} - \hat {\bm {q}} \ \cdot \ {\bm{\kappa}^{(\omega)}} \frac{\partial \left({\bm B}^{(\omega)} {\bm T}^{(\omega)}\right)}{\partial \kappa^{11}_{e}} \right) \nonumber \\ 
\ = & \ - \sum_{\omega \in \Omega_{obj}} \left( \hat {\bm {q}} \ \cdot \ \frac{\partial \bm{\kappa}^{(\omega)}}{\partial \kappa^{11}_{e}} \nabla \bm{T}^{(\omega)} \right) - \sum_{\omega \in \Omega_{obj}} \left(\hat {\bm {q}} \ \cdot \ {\bm{\kappa}^{(\omega)}} \frac{\partial \left({\bm B}^{(\omega)} {\bm L}^{(\omega)} {\bm T}\right)}{\partial \kappa^{11}_{e}} \right)  
\end{align} 
where ${\bm L}^{(\omega)} $ is a matrix associated with the query nodes of the element $\omega$. Finally, we have 

\begin{align} 
\frac{\partial J^{\mathcal{L}}_{ri}}{ \partial \kappa^{11}_{e}} = & \ - \sum_{\omega \in \Omega_{obj}} \left( \hat {\bm {q}} \ \cdot \ \frac{\partial \bm{\kappa}^{(\omega)}}{\partial \kappa^{11}_{e}} \nabla \bm{T}^{(\omega)} \right) - \sum_{\omega \in \Omega_{obj}} \left(\hat {\bm {q}} \ \cdot \ {\bm{\kappa}^{(\omega)}}{\bm B}^{(\omega)} {\bm L}^{(\omega)} \frac{\partial {\bm T}}{\partial \kappa^{11}_{e}} \right) \nonumber \\ 
= & \ - \sum_{\omega \in \Omega_{obj}} \left( \hat {\bm {q}} \ \cdot \ \frac{\partial \bm{\kappa}^{(\omega)}}{\partial \kappa^{11}_{e}} \nabla \bm{T}^{(\omega)} \right) + \sum_{\omega \in \Omega_{obj}} \left(\hat {\bm {q}} \ \cdot \ {\bm{\kappa}^{(\omega)}}{\bm B}^{(\omega)} {\bm L}^{(\omega)} {\bm K}_{T}^{-1}\frac{\partial{{\bm K}_{T}}} {\partial { \kappa^{11}_{e}}} \bm{T} \right) 
\label{riF} 
\end{align}


\subsection{Multiple thermal functionalities} 

To achieve multiple thermal functionalities simultaneously, the sensitivity of the objective function in Eq.~(\ref{MTF}) w.r.t. ${ \kappa^{11}_{e}}$ is written as: 

\begin{align} 
\frac{\partial \left[ \xi_{ck}\frac{J_{ck}}{J_{ck}^{0}} + \xi_{ct} \frac{(J_{ct} -100\%)^2}{(J_{ct}^{0} -100\%)^2}+ \xi_{ri} \frac{J_{ri}}{J_{ri}^{0}} \right]} { \partial \kappa^{11}_{e}} = \frac{\xi_{ck}}{ {J_{ck}^{0}}} \frac{\partial J^{\mathcal{L}}_{ck}}{ \partial \kappa^{11}_{e}} + \frac{\xi_{ct}}{ (J_{ct}^{0} -100\%)^2} \frac{\partial J^{\mathcal{L}}_{ct}}{ \partial \kappa^{11}_{e}} + \frac{\xi_{ri}}{ {J_{ri}^{0}}} \frac{\partial J^{\mathcal{L}}_{ri}}{ \partial \kappa^{11}_{e}}  
\end{align} 
where $ \frac{\partial J^{\mathcal{L}}_{ck}}{ \partial \kappa^{11}_{e}} $, $ \frac{\partial J^{\mathcal{L}}_{ct}}{ \partial \kappa^{11}_{e}} $, and $ \frac{\partial J^{\mathcal{L}}_{ri}}{ \partial \kappa^{11}_{e}} $ are computed by Eqs.~(\ref{eq:st}), (\ref{tctf}), and (\ref{riF}), respectively.

\section{Two-scale thermal property design examples} 

\label{Ep} 

In this section, multiple numerical examples are presented to validate the effectiveness of the proposed data-driven \textit{property} design framework for different thermal functionalities. In all tests, the isotropic material is assumed to have the unit thermal conductivity to directly leverage the constructed database in Section \ref{Model}. At the higher scale for structural finite element analysis, regular meshes using quadrilateral bilinear elements are adopted. All temperatures are in degrees Celsius.


\subsection{Thermal cloak}

\begin{figure}[t!] 
\captionsetup{singlelinecheck = true, justification=justified} 
\centering 
\includegraphics[width=0.96\linewidth]{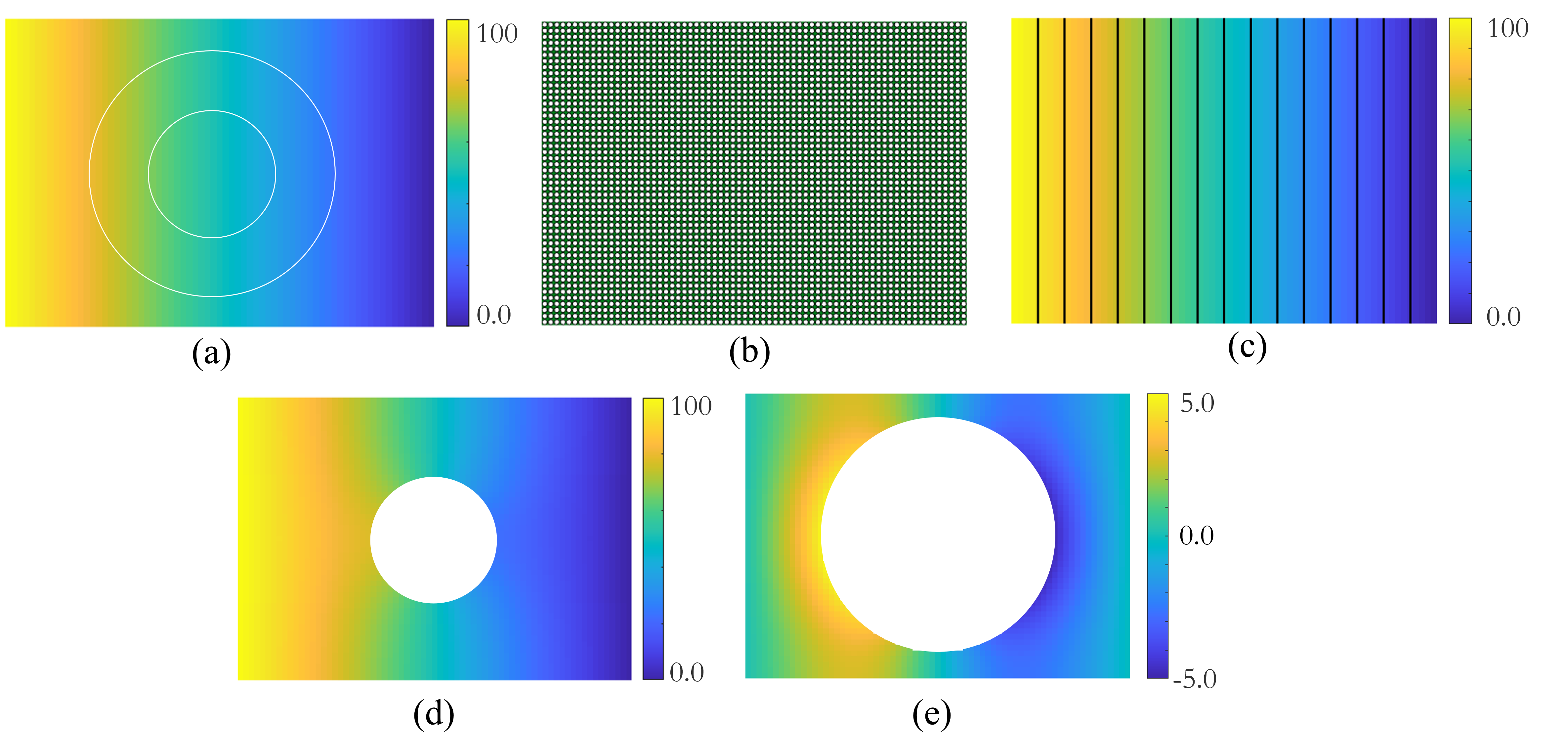} 
\caption{(a) reference temperature profile without any holes and with the indication of the design domain (the region between the two white circles); (b) original structure composed of square RVEs with a central circular hole by assigning only one cell in every structural element; (c) reference temperature profile with the indication of isothermal lines; (d) temperature profile with one central hole; (d) temperature difference between (a) and (d) for the region outside the design domain.} 
\label{fig:CT2} 
\end{figure}

We first consider the domain with the thermal boundary conditions and temperature profile depicted in Fig.~\ref{fig:CT2} (a), where the whole left boundary is the hot source with $100 \ ^\circ \rm C$, and the whole right boundary is the cold source with $0 \ ^\circ \rm C$. The temperature is obtained by assuming that the whole domain is composed of RVEs (at the lower scale) with a 50\% central circular hole. If we use only one RVE to fill in the structural element at the higher scale, the whole structure can be depicted as Fig.~\ref{fig:CT2} (b). In other words, both the design domain (the ring region between the two white circles) and non-design domain (the remaining region) are assumed to be composed of the same RVE for the initial design. The effective thermal conductive property of this RVE is 

\begin{align} 
\boldsymbol \kappa = \begin{bmatrix} 
0.3162 & 0 \\ 
0 & 0.3162 
\end{bmatrix} 
\label{NDD} 
\end{align} 
At the higher structural scale, $75 \times 50$ square elements are used to discretize the whole domain with the unit elemental size. During the structural analysis, the above homogenized effective thermal conductivities are used.

Another efficient way to reflect the temperature profile is to use the isothermal lines. As shown in Fig.~\ref{fig:CT2} (c), each vertical black line is an isothermal line in which the temperature value is the same. Uniform thermal gradient can be observed for this type of boundary conditions. Note that both Figs.~\ref{fig:CT2} (a) and (c) show the temperature profile on the higher structural scale with the scale separation between the RVE and the structure, i.e., not the profile for the structure Fig.~\ref{fig:CT2} (b). Unless otherwise specified, all presented temperature profiles in this work are for the scale separation case. Fig.~\ref{fig:CT2} (d) then shows the temperature profile under the same boundary conditions but with a hole (insulator) inside. Fig.~\ref{fig:CT2} (e) shows the temperature difference between Figs.~\ref{fig:CT2} (a) and (c) for the region outside the design domain.

\begin{figure}[t!] 
\captionsetup{singlelinecheck = true, justification=justified} 
\centering 
\includegraphics[width=0.80\linewidth]{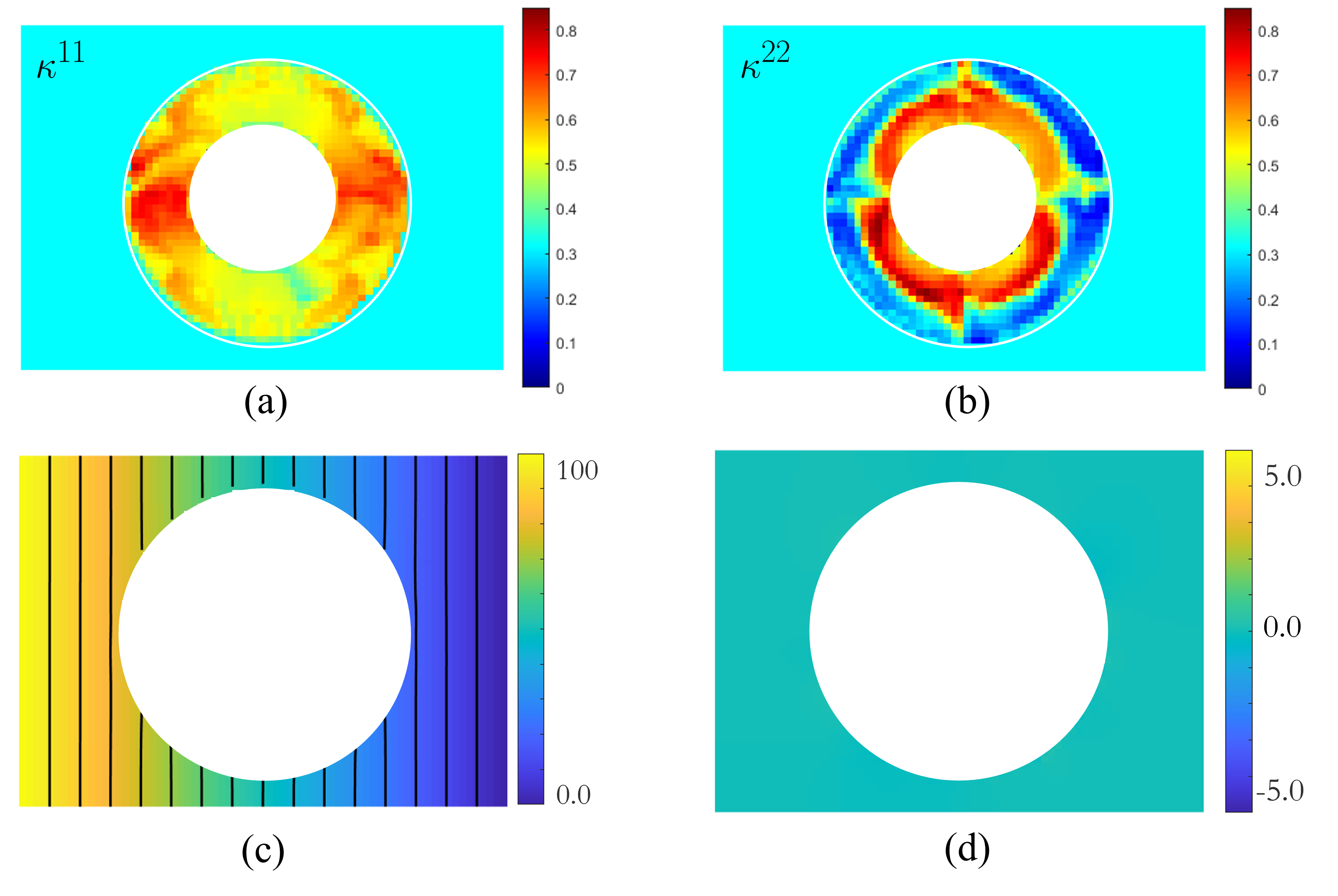} 
\caption{(a) the optimized distribution of the first component of thermal conductivity $\kappa^{11}$; (b) the optimized distribution of the second component of thermal conductivity $\kappa^{22}$; (c) temperature profile with the isothermal lines by using the optimized distribution of thermal conductivities; (d) temperature difference between (c) and Fig.~\ref{fig:CT2} (c) for the region outside the design domain.} 
\label{fig:Result1} 
\end{figure}

After design optimization, the optimized distribution of the two design variables or thermal conductivity components are shown in Figs.~\ref{fig:Result1} (a) and (b). The minimum and maximum values for the optimized first component $\kappa^{11}$ are 0.3250 and 0.7753, and 0.0812 and 0.8404 for the optimized second component $\kappa^{22}$, respectively. Those values are within our database, which includes values from 0 to 1, respectively. Note that the values of the thermal conductivity components for the region outside the design domain are unchanged at 0.3162, for both $\kappa^{11}$ and $\kappa^{22}$. As a result, the new temperature profile with the optimized thermal conductive properties is shown in Figs.~\ref{fig:Result1} (c). We observe that the isothermal lines are exactly the same for the region outside the design domain. Figs.~\ref{fig:Result1} (d) shows the temperature difference between Fig.~\ref{fig:CT2} (c) and Fig.~\ref{fig:Result1} (c), which is nearly 0 everywhere. The value of the objective function (\ref{aa}) for the profile Fig.~\ref{fig:CT2} (c) is as small as 5.5e-4.

\begin{figure}[t!] 
\captionsetup{singlelinecheck = true, justification=justified} 
\centering 
\includegraphics[width=0.96\linewidth]{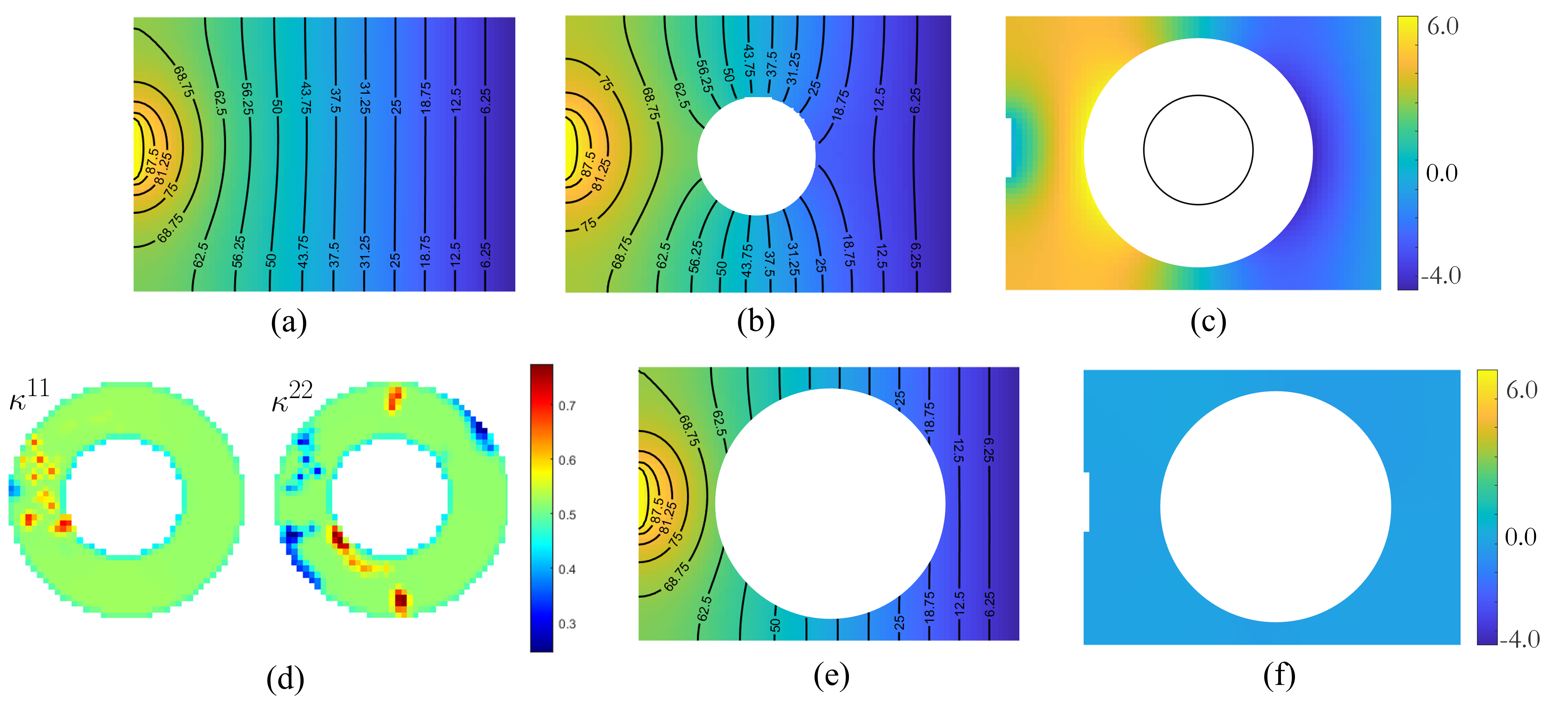} 
\caption{(a) reference temperature profile and isothermal lines for the case of ``inhomogeneous'' thermal gradient; (b) temperature profile and isotherms for ``inhomogeneous'' thermal gradient with a hole; (c) temperature difference between (a) and (b) for the region outside the design domain; (d) two optimized thermal conductive components within design domain; (e) temperature profile and isotherms for ``inhomogeneous''  thermal gradient with a hole and optimized thermal conductive components in the design domain (f) temperature difference between (a) and (e) for the region outside the design domain.} 
\label{fig:Ihom} 
\end{figure}

A challenge in implementing the thermal cloak behavior arises when the reference temperature profile is ``inhomogeneous'' or non-uniform. As shown in Fig.~\ref{fig:Ihom} (a), the heat source is no longer the whole left boundary as Fig.~\ref{fig:CT2} (a), but only a small part of the middle, spanning 10 structural elements. The whole right boundary is still the cold source. The number shown in the isothermal lines is the exact temperature value. Fig.~\ref{fig:Ihom} (b) shows the temperature profile and isothermal lines when there is a hole inside the body. As can be seen, the isothermal line is again severely twisted. Fig.~\ref{fig:Ihom} (c) shows the direct temperature difference between Figs.~\ref{fig:Ihom} (a) and (d) for the domain outside the design domain, in which the temperature difference is quite large. After design optimization, the optimized profile for the two designed components is shown in Fig.~\ref{fig:Ihom} (d). Note that the thermal conductive property of the non-design-domain region is still the same as Eq.~(\ref{NDD}). The minimum and maximum values for the optimized first component in this case are 0.3705 and 0.7134, respectively, and for the optimized second component are 0.2455 and 0.7748, respectively. Again, the values are all within the range of our constructed database. With the optimized thermal conductivity in the design region, the new temperature profile is shown in Fig.~\ref{fig:Ihom} (e). We observe that the isothermal line with a temperature of 68.75 $ \rm {^o} C$, it is now reverting to the left border area instead of the bottom border as in Fig.~\ref{fig:Ihom} (b). Fig.~\ref{fig:Ihom} (f) shows the temperature difference between Figs.~\ref{fig:Ihom} (a) and (e), which is again nearly 0 throughout.

\begin{figure}[t!] 
\captionsetup{singlelinecheck = true, justification=justified} 
\centering 
\includegraphics[width=0.96\linewidth]{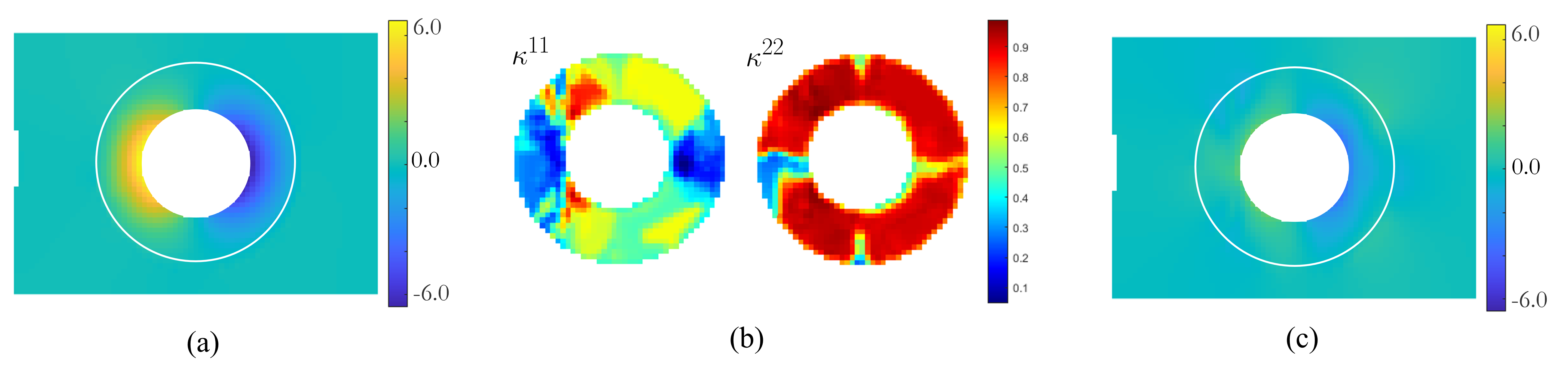} 
\caption{(a) temperature difference between Figs.~\ref{fig:Ihom} (a) and (e) for the region outside and inside the design domain; (b) two optimized thermal conductive components within design domain to cloak everywhere; (c) temperature difference after design optimization.} 
\label{fig:Ihom-case2-2} 
\end{figure}

However, if we plot the temperature difference between Fig.~\ref{fig:Ihom} (a), i.e., the reference solution, and Fig.~\ref{fig:Ihom} (e), i.e., the optimized solution, for the regions not only outside the design domain but also inside the design domain, as shown in Fig.~\ref{fig:Ihom-case2-2} (a), we still find an enormous difference inside the design domain. This is reasonable as for the objective function (\ref{aa}), as the design domain itself is not classified as the area to be cloaked. To solve this issue, which would be a significant challenge using other methods, we then define the Eq.~(\ref{aa-2}) as the new objective function including the ``shield'’ area as the ``cloaking'' area. Note that the design domain is unchanged and retains the same ring area as before. Following a similar design optimization process, the optimized distribution of the two design variables/components is shown in Fig.~\ref{fig:Ihom-case2-2} (b). The maximum values for the first and second optimized components are 0.927 and 0.9890, respectively. The minimum values are 0.0485 and 0.2471, respectively. How to extract the geometry of the high scale structure is detailed in the next section. For the new objective function, Fig.~\ref{fig:Ihom-case2-2} (a) results in a value of 0.536, but only 0.097 for Fig.~\ref{fig:Ihom-case2-2} (c). This difference  is a result of a completely different distribution of the two designed components shown in Fig.~\ref{fig:Ihom-case2-2} (b) versus the optimized components shown in Fig.~\ref{fig:Ihom} (d). Note that the objective function (\ref{aa-2}) from the temperature profile Fig.~\ref{fig:Ihom} (b), i.e., before the optimization, has the value of 5.14.

\subsection{Thermal concentrator}

\begin{figure}[t!] 
\captionsetup{singlelinecheck = true, justification=justified} 
\centering 
\includegraphics[width=0.96\linewidth]{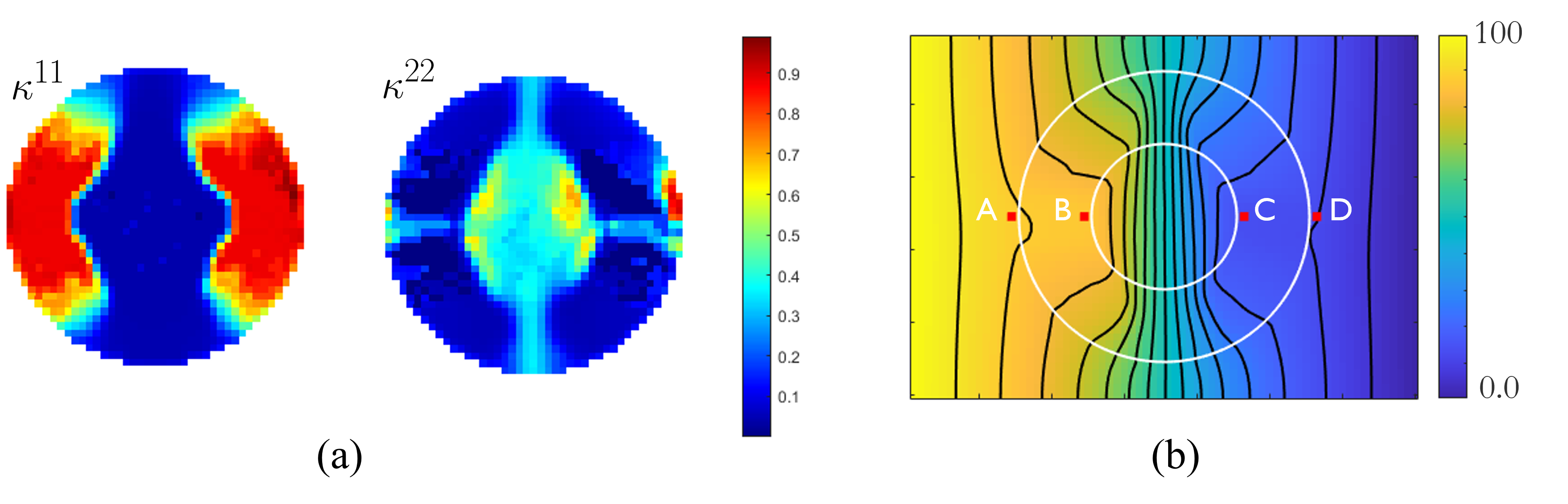} 
\caption{(a) optimized distribution of the two design components for the initial profile Fig.~\ref{fig:CT2} (a); (b) the corresponding temperature profile with the optimized design components (a).} 
\label{tc1} 
\end{figure}

In this subsection, we consider the data-driven two-scale \textit{property} design for the second functionality of thermal concentration. The indicator described in Eq.~(\ref{tct}) is used to account for the degree of the thermal concentration, with values closer to 1 indicating better concentrators. Without any design, the reference solution of the case Fig.~\ref{fig:CT2} (a) has the indicator value of 0.7551. The region inside the large white circle in Fig.~\ref{fig:CT2} (a) serves as the design domain in this subsection. After tailoring the two design components in the design domain, the optimized distribution of the two components is shown in Fig.~\ref{tc1} (a), and the corresponding temperature profile is shown in Fig.~\ref{tc1} (b). It can be seen from Fig.~\ref{tc1} (b) that the heat is highly concentrated in the middle of the whole domain. The value of the indicator is now equal to 0.9653. The maximum values for the first and second components are 0.9910 and 0.9043, respectively, and the minimum values are $2.5\mathrm{e}{-9}$ and $5.0\mathrm{e}{-9}$, respectively.

\begin{figure}[t!] 
\captionsetup{singlelinecheck = true, justification=justified} 
\centering 
\includegraphics[width=0.96\linewidth]{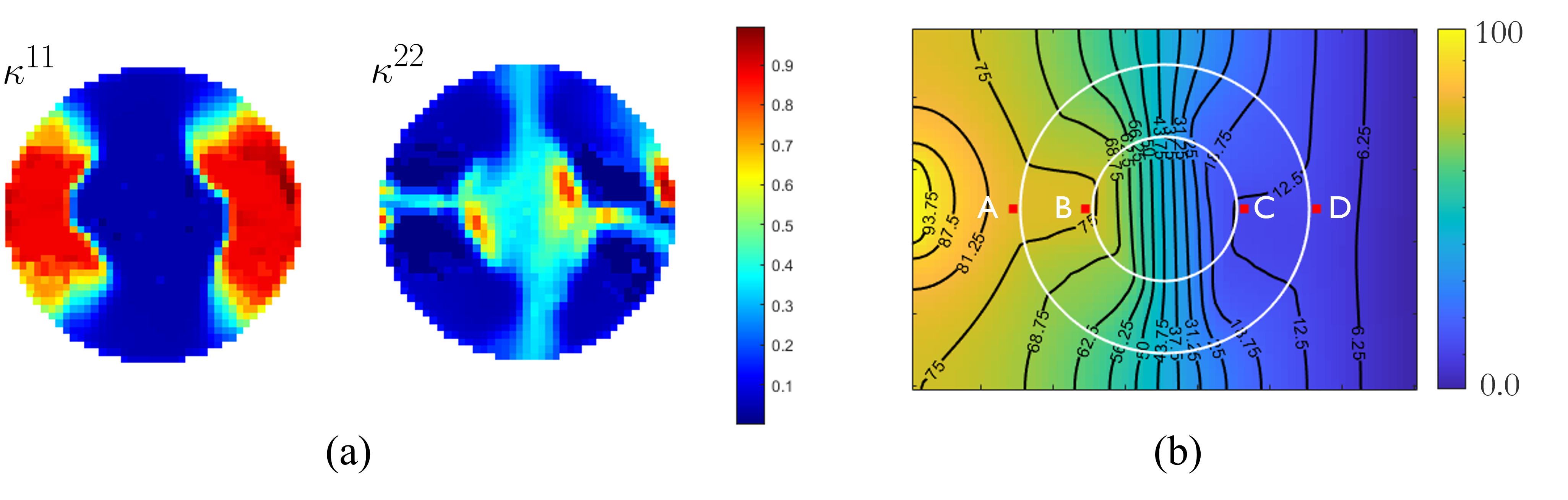} 
\caption{(a) optimized distribution of the two design components for the initial profile Fig.~\ref{fig:Ihom} (a); (b) the corresponding temperature profile with the optimized design components (a).} 
\label{tcc2} 
\end{figure}

\begin{figure}[t!] 
\captionsetup{singlelinecheck = true, justification=justified} 
\centering 
\includegraphics[width=0.96\linewidth]{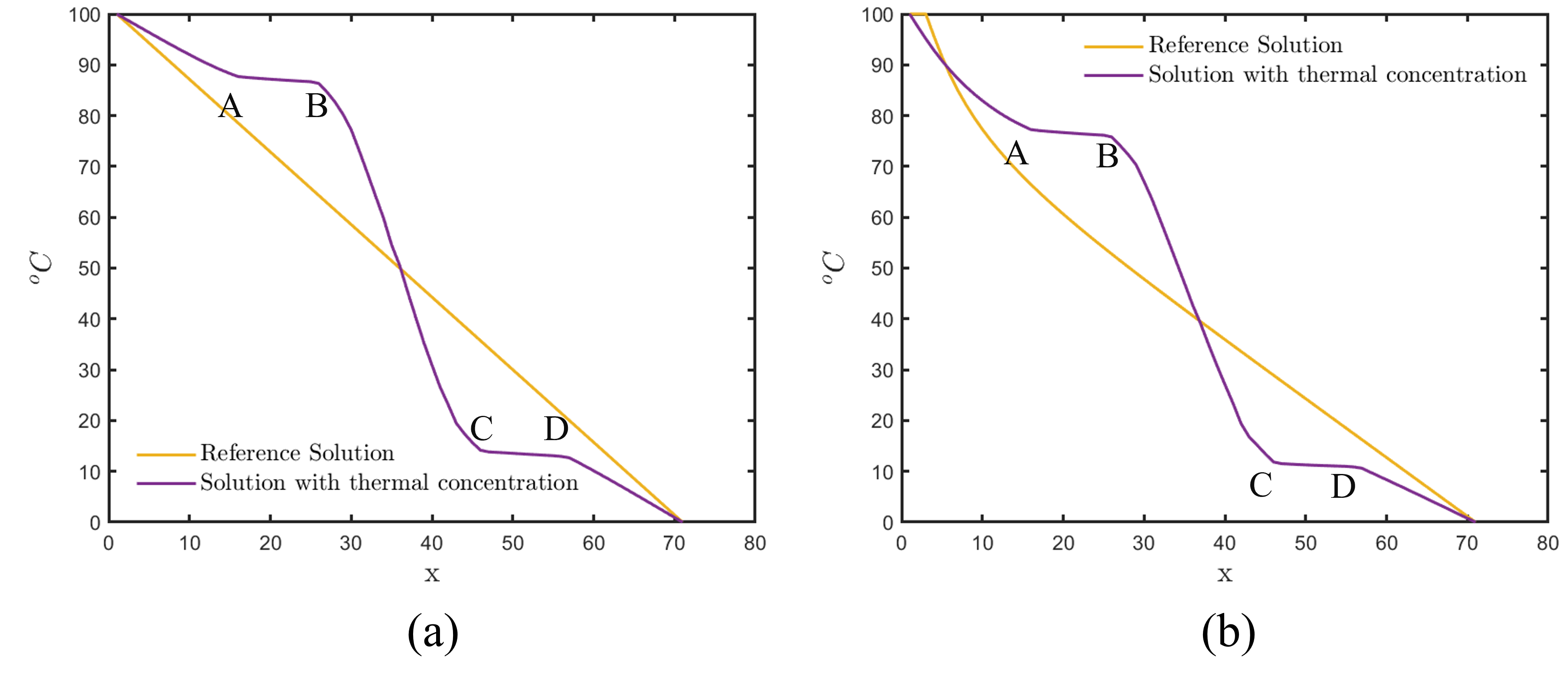} 
\caption{(a) temperature distribution along the line $y = 25$ for Fig.~\ref{tc1}; (b) (a) temperature distribution along the line $y = 25$ for Fig.~\ref{tcc2}.} 
\label{CCcurve} 
\end{figure}

For the reference solution of the non-uniform profile in Fig.~\ref{fig:Ihom} (a), the index value is equal to 0.5238 with poor concentration. After design optimization, the optimized design variable and corresponding temperature profile are shown in Fig.~\ref{tcc2} (a) and (b) respectively. The maximum values for the first and second components are 0.9954 and 0.9542, respectively, and the minimum values are $2.5\mathrm{e}{-9}$ and $5.0\mathrm{e}{-9}$, respectively. The extreme values and overall distribution are surprisingly similar to the previous case, in which the large $\kappa^{11}$ is mainly distributed on the sides, while the large $\kappa^{22}$ is distributed in the middle. From Fig.~\ref{tcc2} (b), the index value is equal to 0.9591 with excellent concentration. Fig.~\ref{CCcurve} shows the temperature distribution along the line $y = 25$ for both the cases of uniform and non-uniform reference profiles.

\begin{figure}[t!] 
\captionsetup{singlelinecheck = true, justification=justified} 
\centering 
\includegraphics[width=0.96\linewidth]{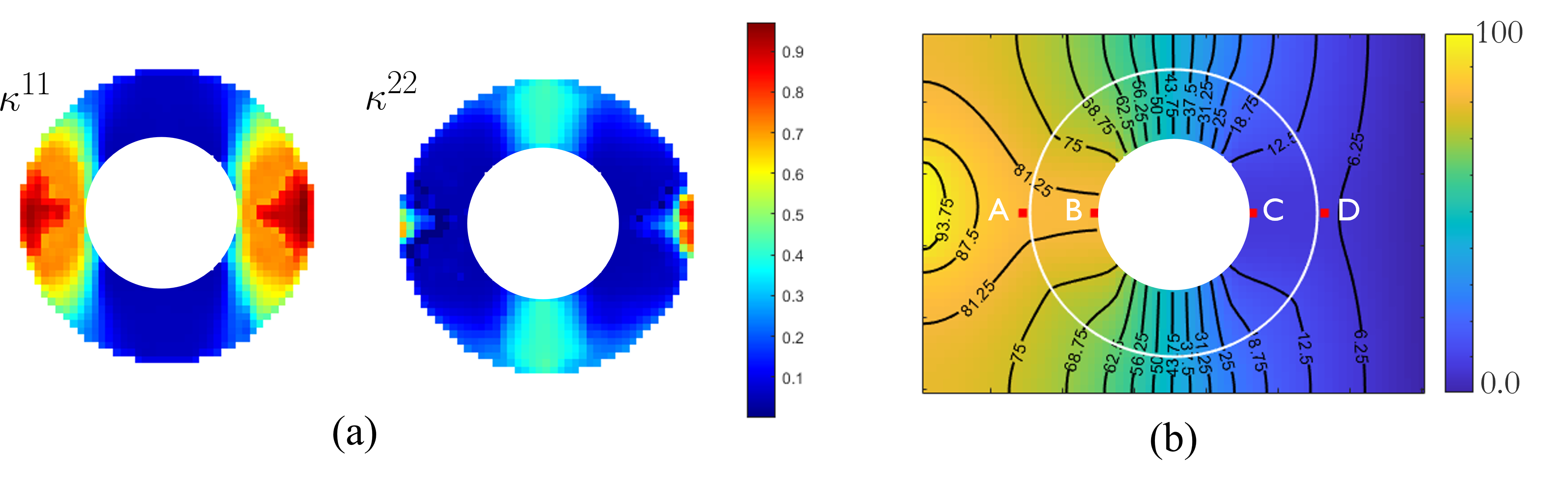} 
\caption{(a) optimized distribution of the two design components for the initial profile Fig.~\ref{fig:Ihom} (b); (b) the corresponding temperature profile with the optimized design components (a).} 
\label{tccc3} 
\end{figure}

Moreover, for the case Fig.~\ref{fig:Ihom} (b) which has the uniform heat source and a hole/insulator inside, the index value is already 0.7910. The design domain in this case is then only the ring region between the two circles. The optimized values for the two design components are shown in Fig.~\ref{tccc3} (a), with the maximum values of 0.9710 and 0.8465, and minimum values of 0.0445 and $5.0\mathrm{e}{-9}$, respectively. For the optimized temperature profile Fig.~\ref{tccc3} (b), the index value is improved to 0.9849.

\subsection{Thermal rotator/inverter} 

\begin{figure}[t!] 
\captionsetup{singlelinecheck = true, justification=justified} 
\centering 
\includegraphics[width=0.96\linewidth]{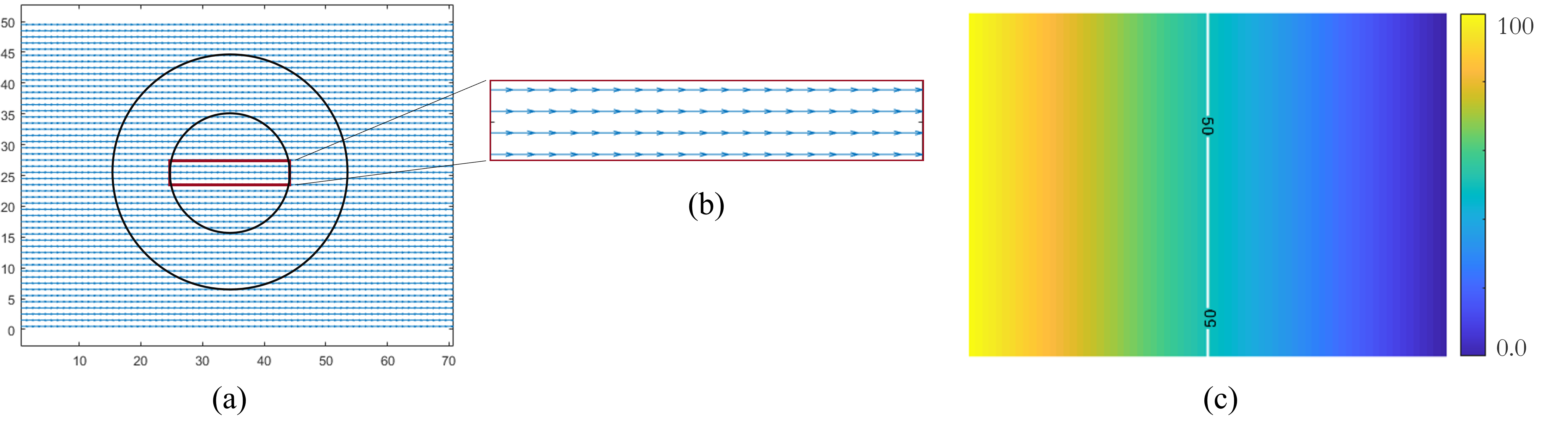} 
\caption{(a) initial heat flux distribution for the temperature profile Fig.~\ref{fig:CT2} (a); (b) heat flow in the target region to be rotated; (c) the corresponding temperature profile with the isotherm at 50 $\rm ^{o} C$.} 
\label{reverse} 
\end{figure}

The design objective of the thermal rotation is considered in this section. The heat flow distribution of the reference temperature profile Fig.~\ref{fig:CT2} (a) is shown in Fig.~\ref{reverse} (a), i.e., from left to right throughout the whole domain. The magnified area shown in Fig.~\ref{reverse} (b) is the target area $\Omega_{obj}$ (see Eq.~(\ref{tri})) for rotating the heat flow. The target area is in the center and is 20 structural elements long and 4 structural elements wide. In this target region, the initial result of Eq.~(\ref{tri}) is 36.1371. Note that the value is positive because the unit vector $\hat {\bm {q}}$ has the same direction as the heat flow in the target region. Fig.~\ref{reverse} (c) replots the temperature profile of Fig.~\ref{fig:CT2} (a) but with the indication of the isotherm at 50 $\rm ^{o} C$, thereby enabling easy comparison with the optimized profiles. The ring region between the two black circles in Fig.~\ref{reverse} (a) serves as the design domain.

\begin{figure}[t!] 
\captionsetup{singlelinecheck = true, justification=justified} 
\centering 
\includegraphics[width=0.96\linewidth]{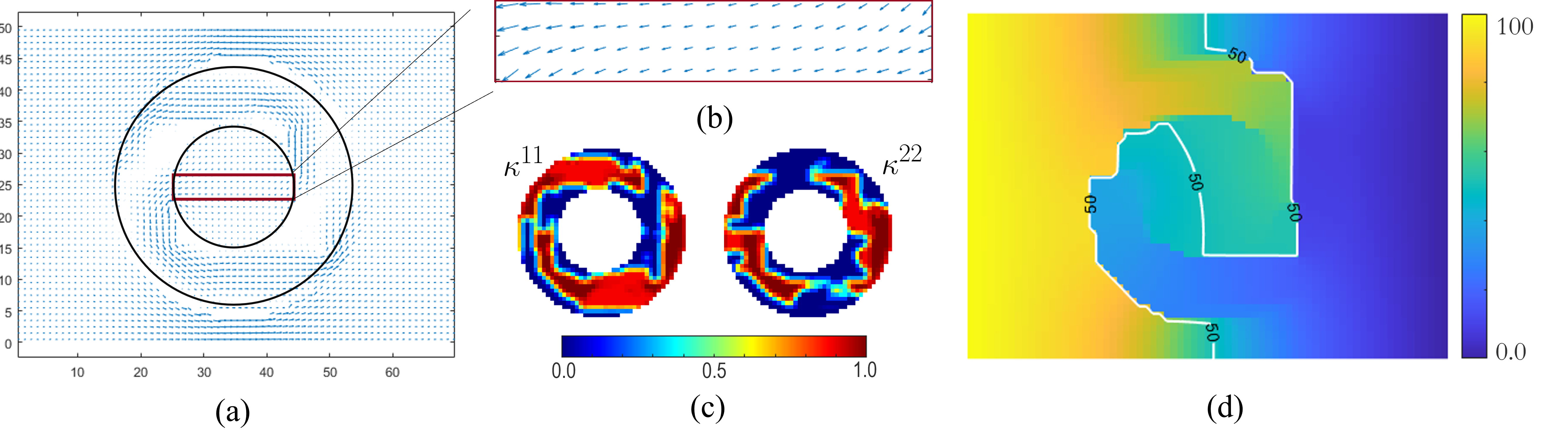} 
\caption{(a) optimized heat flux distributions starting from the initial profile where the thermal conductivity is the same throughout the whole domain (Fig.~\ref{fig:CT2} (a)); (b) heat flow in the target region; (c) the optimized temperature profile with the isotherm at 50 $\rm ^{o} C$.} 
\label{reverse2}  
\end{figure}

After design iterations, the optimized heat flow is shown in Fig.~\ref{reverse2} (a). The heat flow in the same enlarged region is shown in Fig.~\ref{reverse2} (b), where the heat flux is totally reversed. The resulting value for Eq.~(\ref{tri}) is -20.4483, which is negative because of the heat inverter. Again, this is because the optimized thermal conductive distribution in the ring design domain, as shown in Fig.~\ref{reverse2} (c). The maximum values for the optimized two components are both equal to 1.0, while the minimum values for the first ($\kappa^{11}$) and second ($\kappa^{22}$) components are $1.7\mathrm{e}{-9}$ and $2.3\mathrm{e}{-9}$, respectively. Fig.~\ref{reverse2} (c) shows the corresponding temperature profile. Compared to Fig.~\ref{reverse} (c), the isotherm at 50 $\rm ^{o} C$ is severely distorted, and it can be clearly seen that the temperature of the local right area is higher than the left area.

\begin{figure}[t!] 
\captionsetup{singlelinecheck = true, justification=justified} 
\centering 
\includegraphics[width=0.96\linewidth]{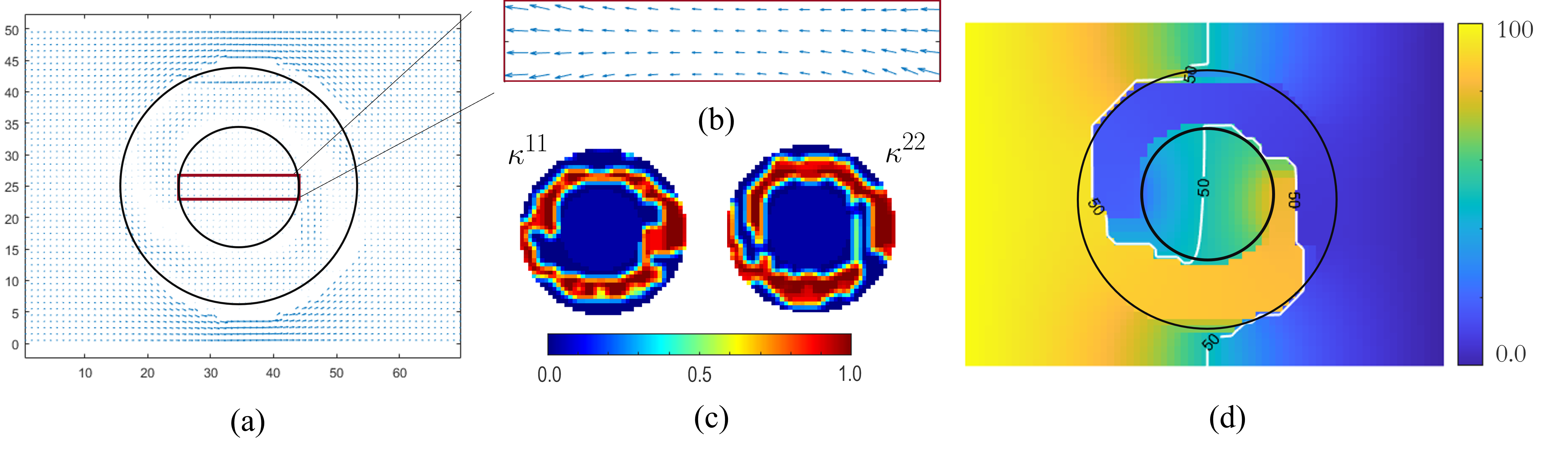} 
\caption{(a) optimized heat flux distribution starting from the initial profile where the thermal conductivity inside the small black circle is 10 times smaller than the outside; (b) heat flow in the target region; (c) the optimized temperature profile with the isotherm at 50 $\rm ^{o} C$.} 
\label{reverse3} 
\end{figure}

Instead of making the thermal conductivity inside the small black circle the same as the conductivity outside the large circle (see Fig.~\ref{reverse} (a) for example), we consider the conductivity inside the small circle to be 10 times weaker, i.e., $\kappa = 0.0316$. The design domain and target region to rotate the heat flow is the same as before. After a similar optimization process, the optimized heat flow distribution is shown in Fig.~\ref{reverse3} (a), and the heat flow in the same enlarged region is shown in Fig.~\ref{reverse3} (b). In this case, the objective function is reduced from positive (6.3168) to negative (-7.7632). Although a similar heat flow distribution can be seen in Fig.~\ref{reverse2} (b) and Fig.~\ref{reverse3} (b), it is interesting to note that, for Fig.~\ref{reverse2} (b), the heat flow is first conducted from left to right \textit{above} the ring area, and then from right to left in the enlarged area. By contrast, for Fig.~\ref{reverse3} (b), the heat flow is first conducted from left to right \textit{under} the ring area, and then from right to left in the enlarged area. Both rotation directions can achieve the performance of thermal reversal in the local area. This finding is further illustrated in Figs.~\ref{reverse3} (c) and (d). In Fig.~\ref{reverse3} (c), the lower left part has a large number of high-value thermal conductivity distributions. The maximum values for the two optimized components both reached 1.0, and the minimum value was again almost 0. In Fig.~\ref{reverse3} (d), the bottom of the ring has a much higher temperature than the top.

\subsection{Multiple functionalities} 

With the above design optimization model established, it is convenient for us to pursue multiple thermal functionalities simultaneously in one design, as formulated in Eq.~\ref{MTF}. By studying firstly thermal cloaking and concentration, we select $\xi_{ck} = 1.5$, $\xi_{ct} = 0.5$, and $\xi_{ri} = 0$. Similar to the problem settings for Fig.~\ref{reverse3}, the inside small circle is filled with a material possessing a thermal conductivity $\kappa = 0.0316$. The design domain is still the ring region. After design optimization, the obtained property distribution is shown in Fig.~\ref{MF}. The maximum values for the optimized two components ($\kappa^{11}$ and $\kappa^{22}$) are equal to 0.9813 and 0.9076, while the minimum values are 0.0031 and $1.1\mathrm{e}{-9}$, respectively. The objective function for thermal cloak is reduced from the original value of 88.17 to 0.26, while the heat concentration index is increased from 0.9372 to 0.9996, demonstrating excellent concentration and cloaking phenomena.

\begin{figure}[t!]  
\captionsetup{singlelinecheck = true, justification=justified}  
\centering  
\includegraphics[width=0.96\linewidth]{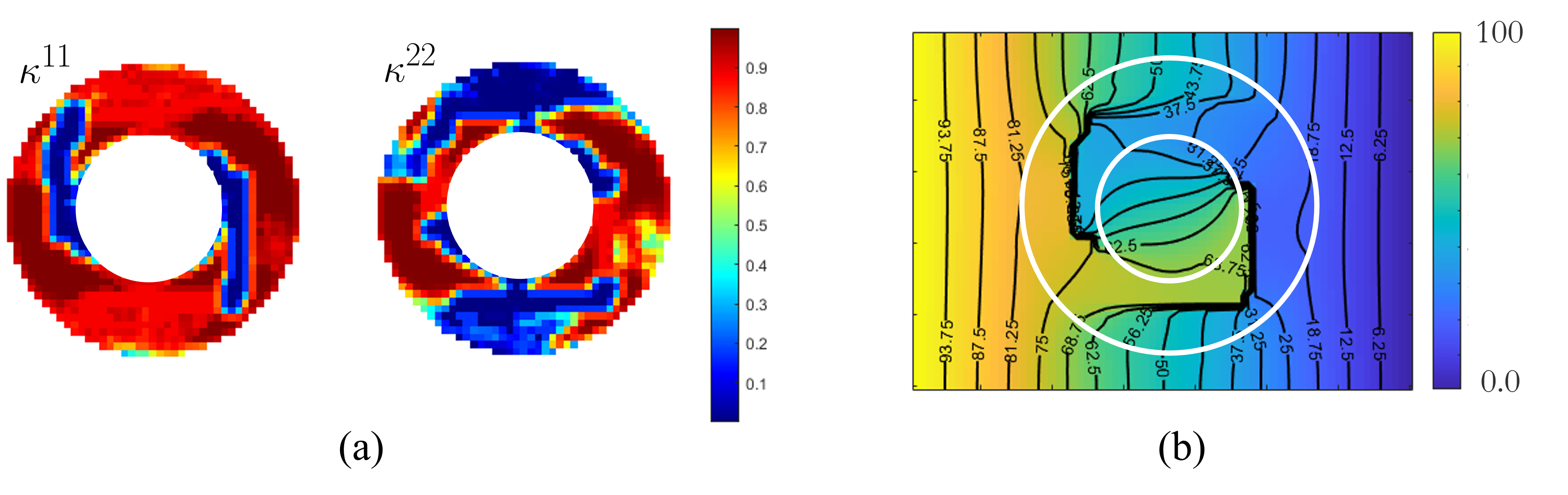}  
\caption{(a) optimized distribution of the two design components for thermal cloak and rotation/inverse simultaneously; (b) the corresponding temperature profile with the optimized design components (a).}  
\label{MF2}  
\end{figure}

For simultaneous cloaking and rotation, the initial condition is set the same as above, and we select $\xi_{ck} = 1.5$, $\xi_{ct} = 0$, and $\xi_{ri} = 5$. After design optimization, the obtained property distribution is shown in Fig.~\ref{MF2}. The objective function for the thermal cloak is reduced from original 88.17 to final 24.40, while it is decreased from 6.3168 to -3.0463 for thermal rotation. The maximum values for the optimized two components are both equal to 1, while the minimum values are $5.0\mathrm{e}{-9}$ and $3.1\mathrm{e}{-9}$, respectively. Again, both objective functions for the required functionalities are simultaneously improved after the \textit{property} design.

\begin{figure}[t!]  
\captionsetup{singlelinecheck = true, justification=justified}  
\centering  
\includegraphics[width=0.96\linewidth]{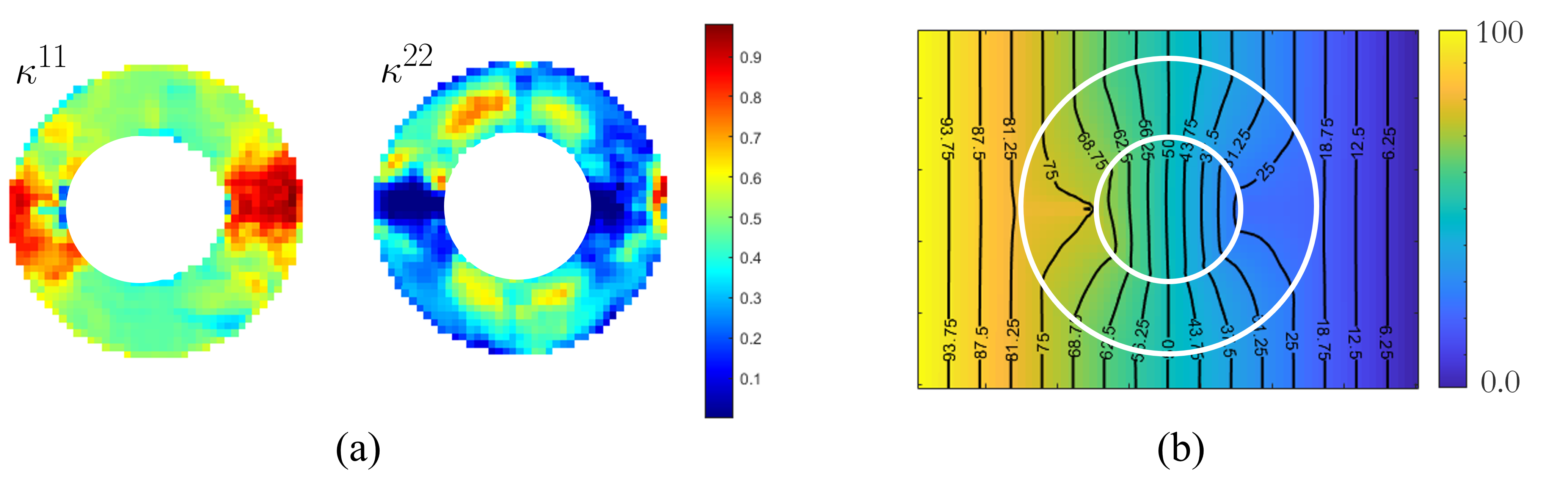}  
\caption{(a) optimized distribution of the two design components for thermal cloak and concentration simultaneously; (b) the corresponding temperature profile with the optimized design components (a).}  
\label{MF}  
\end{figure}

\section{Extract unit cells and structures}

After obtaining the optimized property distribution, the next step is to find unit cells and assemble the final heterostructures. There will be a mapping between the thermal conductivity ($\kappa^{11}$ and $\kappa^{22}$) and the RVE architecture in order to identify unit cells and assemble the final heterostructures. During the mapping and assembling, two criteria generally need to be considered. The first is that the RVE found needs to have optimized thermal conductivity. The second is that there should be good connectivity between adjacent RVEs. The second criterion is automatically satisfied by the adopted database, so we need only to consider the first. In other words, we need to search the database to find different RVEs to meet their optimized thermal conductivities in different local elements. Note that for illustrative purposes, only one RVE is used to fill a structural element.

\begin{figure}[t!] 
\captionsetup{singlelinecheck = true, justification=justified} 
\centering 
\includegraphics[width=0.9\linewidth]{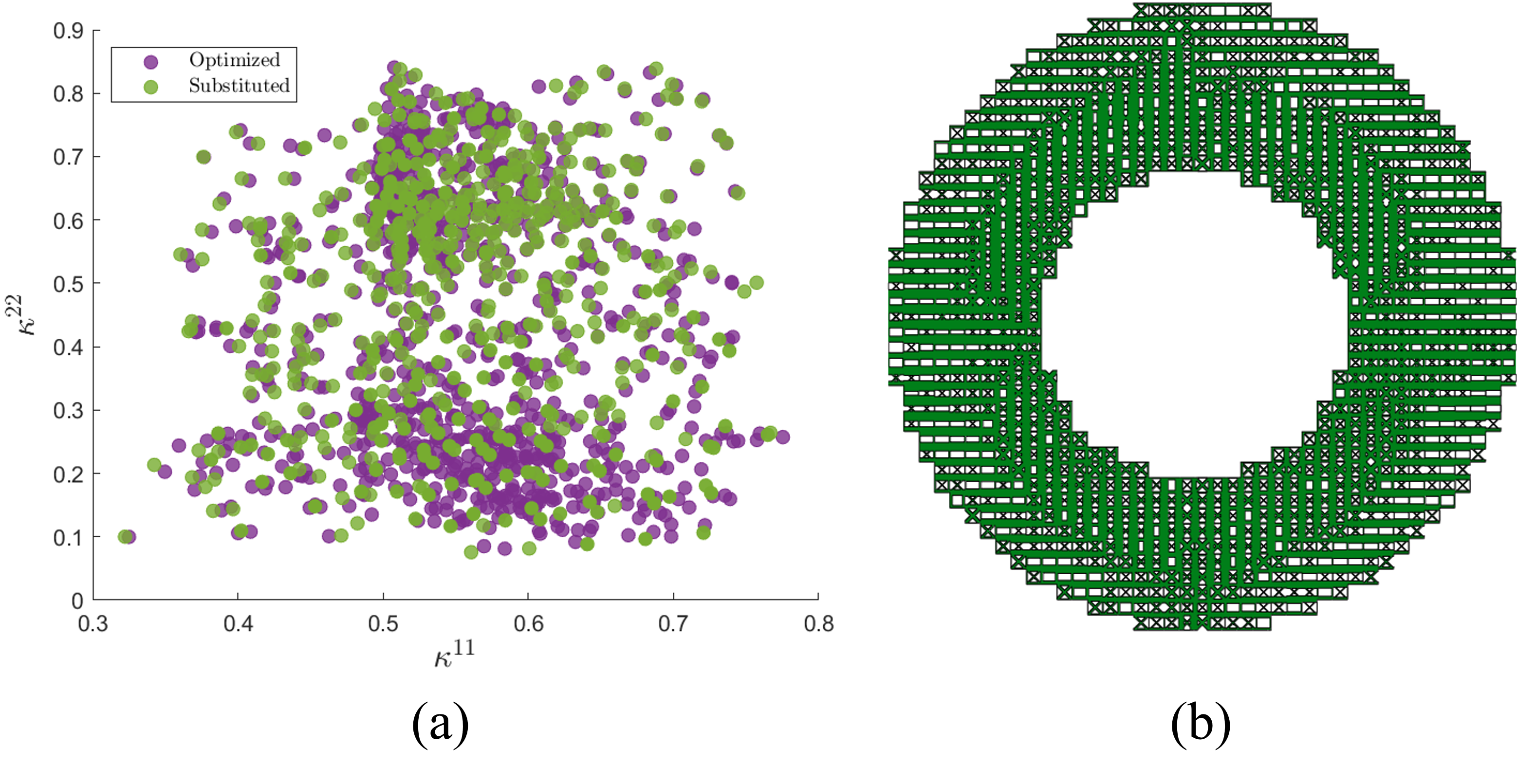} 
\caption{(a) scatter plot for the optimized property from Fig.~\ref{fig:Result1} and the substituted property from the constructed database; (b) the assembled structure with one substituted cell in each structural element.} 
\label{S1} 
\end{figure}

Three structures are to be assembled for the thermal cloak based on the optimized property distribution in Figs.~\ref{fig:Result1}, \ref{fig:Ihom}, and \ref{fig:Ihom-case2-2}. For Fig.~\ref{fig:Result1}, a scatter plot for the optimized two components $\kappa^{11}$ and $\kappa^{22}$ shown in purple in Fig.~\ref{S1} (a). Through searching the database, we find the closest one to every purple dot and display it in green in Fig.~\ref{S1} (a). We obtain the substitution by calculating the difference between the two optimal values of each structural element and the whole database, taking the absolute value, summing, and then designating the smallest one as the substituted one. Both the \textit{mean squared error (MSE)} and the \textit{coefficient of determination ($R^2$)} are computed to measure the quality of the database substitution. The \textit{mean squared error (MSE)} is computed by:

\begin{equation} 
MSE = \frac{1}{N_e} \sum_{e=1}^{N_e} (\kappa_{e}^{Opt} - {\kappa_{e}^{Sub}} )^2 
\label{MSE} 
\end{equation} 
where $N_{e}$ is the number of structural elements in the ring design domain. $\kappa_{e}^{Opt} $ is a column vector containing both the optimized $\kappa^{11}$ and $\kappa^{22}$, and $\kappa_{e}^{Sub} $ is the corresponding vector with the substituted $\kappa^{11}$ and $\kappa^{22}$ from the constructed database. The resulted value is $4.0\mathrm{e}{-5}$ for Fig.~\ref{S1} (a) indicating a excellent match between these two.

The \textit{coefficient of determination ($R^2$)} is computed by: 

\begin{equation} 
R^2 = 1 - \frac{ \sum_{e=1}^{N_e} (\kappa_{e}^{Opt} - {\kappa_{e}^{Sub}} )^2 }{ \sum_{e=1}^{N_e} (\kappa_{e}^{Opt} - {\bar{\kappa}}_{e}^{Opt} )^2 } 
\label{RSquare} 
\end{equation} 
where ${\bar{\kappa}}_{e}^{Opt}$ is the mean of the optimized data. In the best case, the modeled values exactly match the observed values, which results in ${ R^{2}=1}$. Not surprisingly, our $R^2$ has a value of 0.9986 for Fig.~\ref{S1} (a). Fig.~\ref{S1} (b) shows the assembled structural geometry when each designed element is filled in with one substituted cell. Note that in the database, there is a one-to-one mapping between the thermal conduction property and the RVE architecture, so when the substituted property is determined, the structural geometry is unique. Admittedly, different RVE architectures may have the same equivalent thermal conductivity. In this case, we simply select the structural geometry that ranks first in the database. This convenience depends on all the architectures in our database being well connected.

\begin{figure}[t!] 
\captionsetup{singlelinecheck = true, justification=justified} 
\centering 
\includegraphics[width=0.9\linewidth]{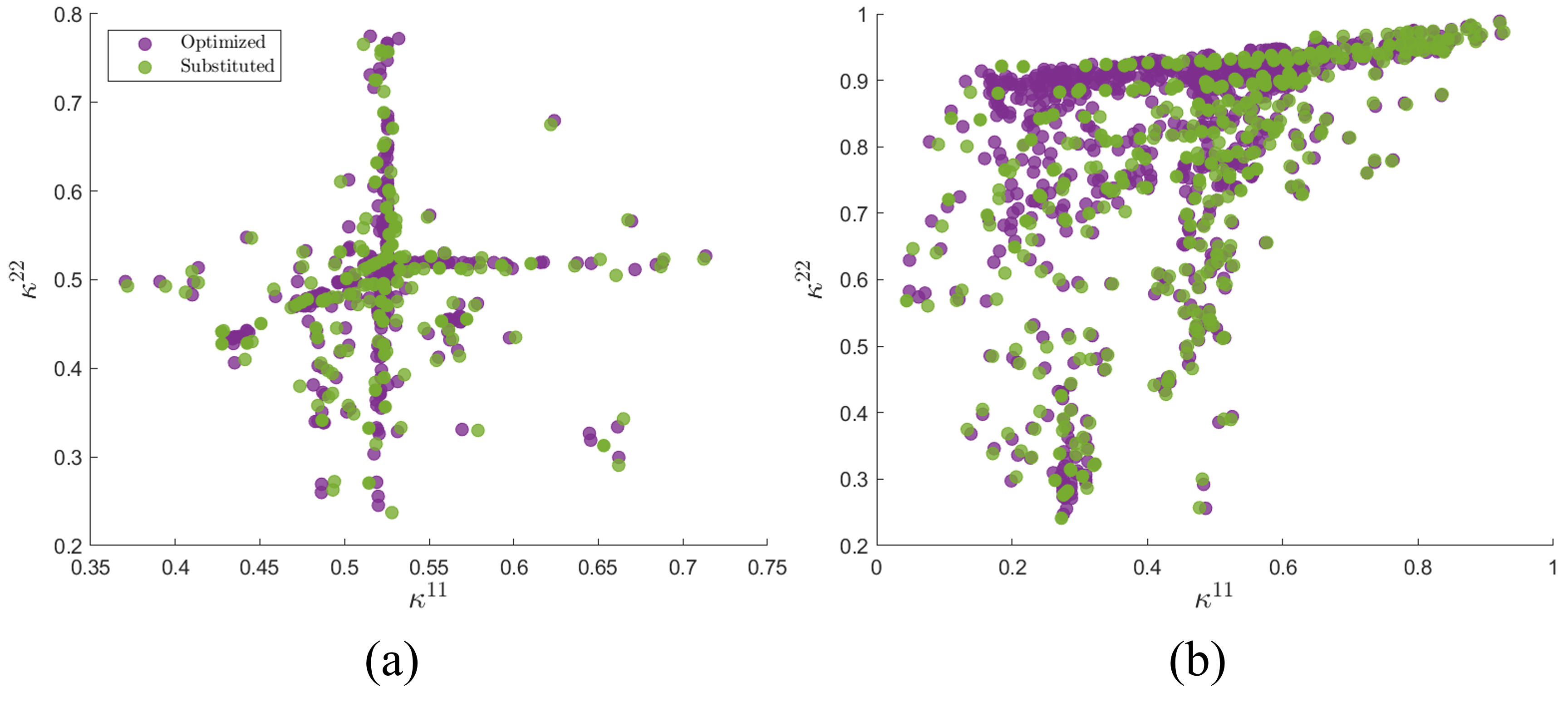} 
\caption{(a) scatter plot for the optimized property from Fig.~\ref{fig:Ihom} and the substituted property from the constructed database; (b) scatter plot for the optimized property from Fig.~\ref{fig:Ihom-case2-2} and the substituted property from the constructed database.} 
\label{S3} 
\end{figure}

Scatter plots of both optimized and substituted values for the optimized cases Figs.~\ref{fig:Ihom} and \ref{fig:Ihom-case2-2} are shown in Figs.~\ref{S3} (a) and (b), respectively. The \textit{MSE} and \textit{$R^2$} for the case Fig.~\ref{S3} (a) are $9.3\mathrm{e}{-6}$ and 0.9947, and for Fig.~\ref{S3} (b) are $7.7\mathrm{e}{-5}$ and 0.9987, respectively. All of which illustrate an excellent substitution.

\begin{figure}[t!] 
\captionsetup{singlelinecheck = true, justification=justified} 
\centering 
\includegraphics[width=0.9\linewidth]{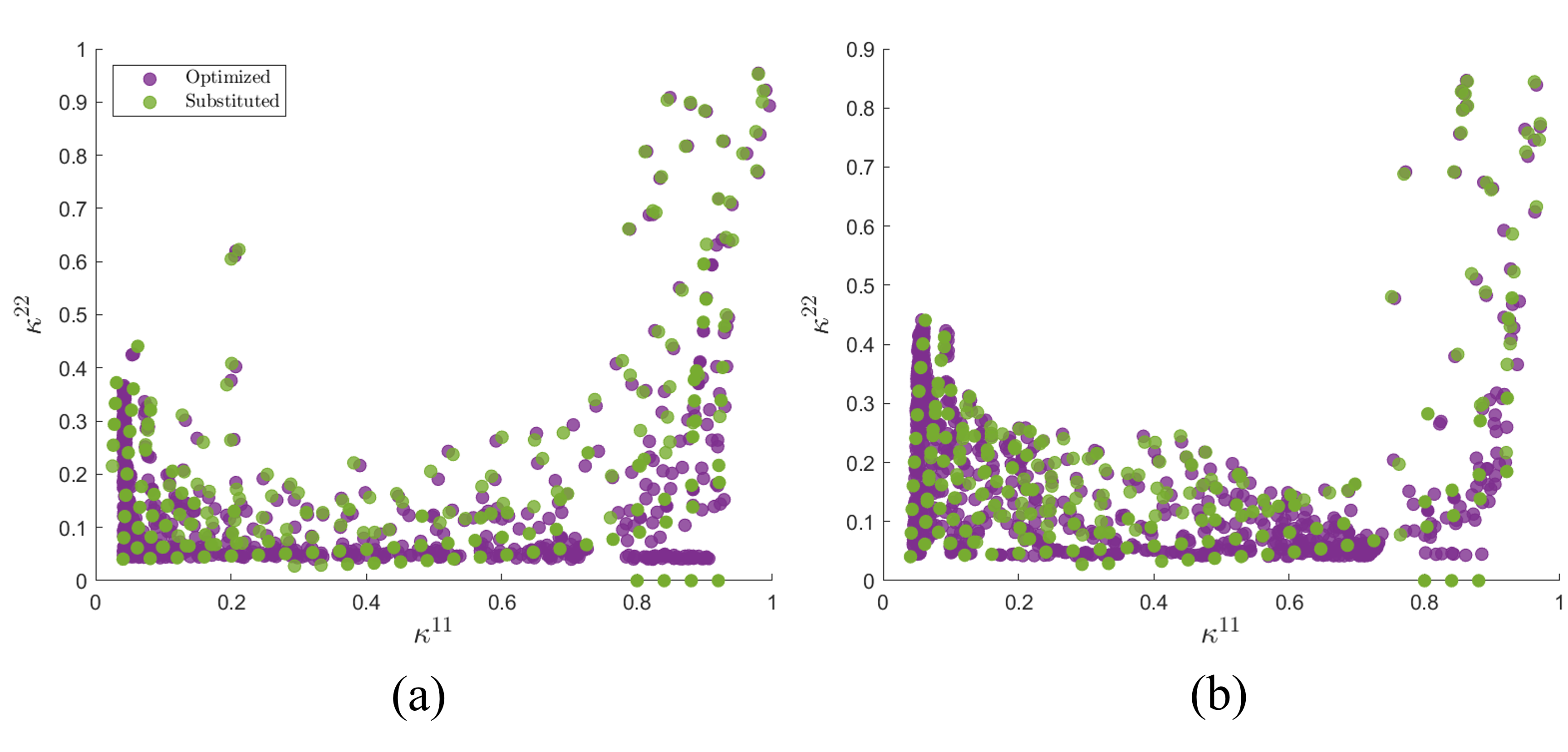} 
\caption{(a) scatter plot for the optimized property from Fig.~\ref{tcc2} and the substituted property from the constructed database; (b) scatter plot for the optimized property from Fig.~\ref{tccc3} and the substituted property from the constructed database.} 
\label{S4} 
\end{figure}

For thermal concentrator, scatter plots of both optimized and substituted values for the optimized cases Figs.~\ref{tcc2} and \ref{tccc3} are shown in Figs.~\ref{S4} (a) and (b), respectively. The \textit{MSE} and \textit{$R^2$} for Fig.~\ref{S4} (a) are $1.2\mathrm{e}{-3}$ and 0.9989, and for Fig.~\ref{S4} (b) are $1.1\mathrm{e}{-4}$ and 0.9984, respectively.

\begin{figure}[t!] 
\captionsetup{singlelinecheck = true, justification=justified} 
\centering 
\includegraphics[width=0.9\linewidth]{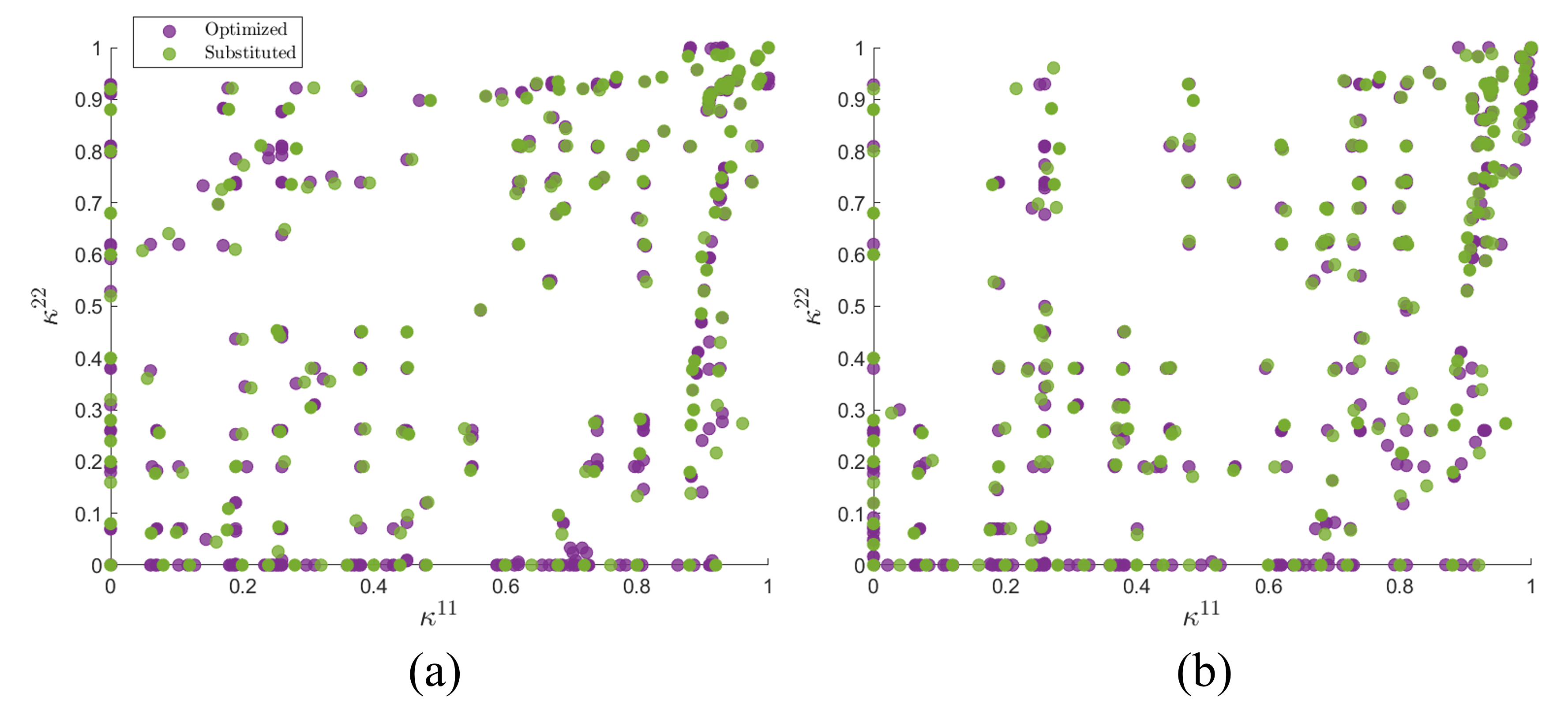} 
\caption{(a) scatter plot for the optimized property from Fig.~\ref{reverse2} and the substituted property from the constructed database; (b) scatter plot for the optimized property from Fig.~\ref{reverse3} and the substituted property from the constructed database.} 
\label{S5} 
\end{figure}

Correspondingly, the scatter plots for thermal rotation in the cases of Figs.~\ref{reverse2} and \ref{reverse3} are shown in Figs.~\ref{S5} (a) and (b), respectively. The \textit{MSE} and \textit{$R^2$} for Fig.~\ref{S5} (a) are $1.1\mathrm{e}{-4}$ and 0.9984, and for Fig.~\ref{S5} (b) are $3.7\mathrm{e}{-5}$ and 0.9998, respectively. Again, all of which illustrate an excellent substitution.

\begin{figure}[t!]  
\captionsetup{singlelinecheck = true, justification=justified}  
\centering  
\includegraphics[width=0.9\linewidth]{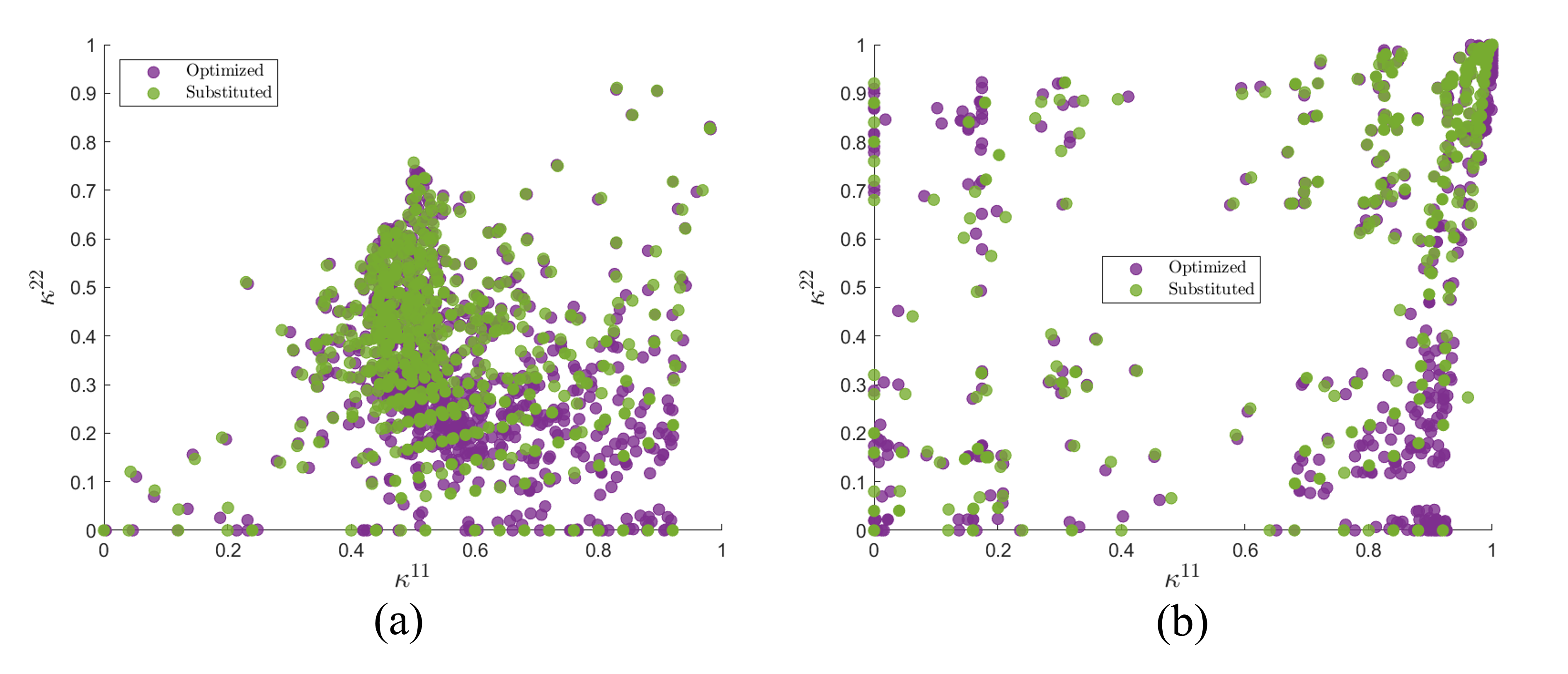}  
\caption{(a) scatter plot of the optimized and substituted properties for thermal cloak and concentration; (b) scatter plot of the optimized and substituted properties for thermal cloak and rotation.}  
\label{MFerror}  
\end{figure}

For the two case studies with multiple objective functions, the scatter plots are shown in Fig.~\ref{MFerror} (a) and (b). For thermal cloak and concentration in (a), the \textit{MSE} and \textit{$R^2$} are $7.1\mathrm{e}{-5}$ and 0.9981, respectively. For thermal cloak and rotation functionalities in (b), the \textit{MSE} and \textit{$R^2$} are $7.0\mathrm{e}{-5}$ and 0.9995, respectively, which demonstrate excellent matching between the optimized and substituted ones for optimizing multiple objective functions. 

\section{Conclusions} 
\label{CONS}

This study presents a novel approach to multiscale data-driven design for achieving multiple thermal functionalities, at the structure scale, including thermal cloaks, thermal concentrators, thermal rotators/inverters, and their combinations. By constructing a comprehensive database containing various unit cell architectures and their corresponding homogenized thermal conductivity, we achieve the multiple macroscopic thermal functionalities, by tailoring the underlying unit cell architecture and conductivity, and using only a single isotropic material. The design optimization framework enables us to tackle challenges that are difficult to solve using other strategies, such as cloaking the ``shielding" area and manipulating non-uniform temperature profiles. Additionally, the approach of optimizing the components of the homogenized thermal conductivity instead of the topology associated with each unit cell significantly reduces the design dimensionality. 

Geometric extraction is performed to compare the optimized homogenized conductivity with the one extracted from the constructed database for both single and multiple functionalities in one design. The results show an excellent match between the two, with a mean squared error as low as 9.3e-6 and a coefficient of determination as high as 0.9998. These advanced thermal metamaterials pave the way for multiscale components with a broad range of heat transfer applications. In conclusion, this work shows the effectiveness of the proposed multiscale data-driven design approach and its potential for creating advanced thermal metamaterials with tailored functionalities.

\section*{Acknowledgements}
This work is supported by NSF CSSI program (Grant No. OAC 1835782).
\bibliographystyle{elsarticle-num}
\bibliography{TDOHM}

\end{document}